\documentclass[14pt]{book}
\usepackage[sectionbib]{natbib}
\usepackage{array,epsfig,fancyheadings,rotating}
\usepackage{amsmath}
\usepackage{amssymb}
\usepackage{amsfonts}
\usepackage{multirow}
\usepackage{amsthm}
\usepackage{lineno,hyperref}
\usepackage{amsfonts}
\usepackage{amssymb}
\usepackage{amsmath}
\usepackage{color}
\usepackage{graphicx}
\usepackage{float}
\usepackage{subfigure}
\usepackage{indentfirst}
\usepackage{bm}
\usepackage{graphics}
\usepackage{epsfig,amssymb,latexsym,verbatim}
\usepackage{array}
\usepackage{booktabs}
\usepackage{multicol}
\usepackage{multirow}
\usepackage[english]{babel}
\usepackage{epstopdf}
\usepackage{amsthm}
\usepackage{natbib}
\usepackage{mathrsfs}
\usepackage[mathscr]{eucal}

\floatstyle{ruled}
\newfloat{algorithm}{tbp}{loa}
\floatname{algorithm}{Algorithm}
\newtheorem{theorem}{Theorem}
\newtheorem{lemma}{Lemma}
\newtheorem{corollary}{Corollary}

\theoremstyle{definition}

\input{tcilatex}
\def\theequation{\arabic{equation}}
\def\thelemma{\arabic{lemma}}

\renewcommand{\baselinestretch}{2}

\begin{document}

\markright{
}
\markboth{\hfill{\footnotesize\rm LI CAI, LIJIE GU, QIHUA WANG, AND SUOJIN WANG}\hfill}
{\hfill {\footnotesize\rm SIMULTANEOUS CONFIDENCE BANDS FOR MISSING COVARIATES DATA } \hfill}
\renewcommand{\thefootnote}{} $\ $


\fontsize{12}{14pt plus.8pt minus .6pt}\selectfont
\vspace{0.8pc}
\centerline{\large{\bf Simultaneous confidence bands for nonparametric regression   }} \vspace{2pt}
\centerline{\large{\bf with partially missing covariates }} \vspace{.4cm}

\centerline{Li Cai$^1$, Lijie Gu$^2$ , Qihua Wang$^1$, and Suojin Wang
$^3$} \vspace{.4cm}
\centerline{\it $^1$Zhejiang Gongshang University,
$^2$Soochow University,  and
$^3$Texas A\&M University}

\selectfont


\begin{quotation}
\noindent \textit{Abstract:} In this paper, we consider a weighted local linear estimator based on the inverse selection probability for nonparametric regression with missing covariates at random. The asymptotic distribution of the maximal deviation between the estimator and the true regression function is derived and an asymptotically accurate simultaneous confidence band is constructed. The estimator for the regression function is shown to be oracally efficient in the sense that it is uniformly indistinguishable from that when the selection probabilities are known. Finite sample performance is examined via simulation studies which support our asymptotic theory. The proposed method is demonstrated via an analysis of a data set from the Canada 2010/2011 Youth Student Survey.

\vspace{9pt} \noindent \textit{Key words and phrases:} Brownian motion,
Maximal deviation, Simultaneous confidence
band, Weighted local linear regression.
\end{quotation}

\fontsize{12}{14pt plus.8pt minus .6pt}\selectfont
\setcounter{chapter}{1} \setcounter{equation}{0}

\noindent {\large \textbf{1. Introduction\label{SEC:Introduction}} }

In nonparametric data analysis, one important problem is to detect the
global shape of unknown curves or to test whether these curves follow some
specific functional forms that describe the overall trend of the regression
relationship. Many researchers have attempted to solve this problem by
constructing nonparametric simultaneous confidence bands ({SCBs}) as a vital
tool of global inference for unknown curves; see Johnston (1982), Zhou et
al. (1998), Fan and Zhang (2000), Claeskens and Van Keilegom (2003), Zhao
and Wu (2008), Cao et al. (2012), Cai et al. (2014), Cao et al. (2016), Cai
et al. (2019a) for the related theory and applications.

Consider the common situation where observations $\left(
X_{i},Y_{i},\varepsilon _{i}\right) _{i=1}^{n}$ are independent and
identically distributed (i.i.d.) copies of $\left( X,Y,\varepsilon \right) $
satisfying the following nonparametric regression model
\begin{equation}
Y=m\left( X\right) +\varepsilon ,  \label{DEF:nonpara model}
\end{equation}%
where $\limfunc{E}\left( \varepsilon |X\right) =0$, $\limfunc{var}\left(
\varepsilon |X\right) =\sigma ^{2}\left( X\right) $, and the mean function $%
m\left( \cdot\right) $ and the variance function $\sigma ^{2}\left( \cdot
\right) $ defined on a compact interval $\left[ a,b\right] $ are unknown. In
order to construct an asymptotically accurate SCB for the mean function $%
m\left( x\right) $, one requires to find a bound $L_{\alpha }$ such that $%
\lim_{n\rightarrow \infty }P\left( \sup_{x\in \left[ a,b\right] }\left\vert
\hat{m}\left( x\right) \right . \right. $ $\left . \left . -m\left( x\right) \right\vert \leq L_{\alpha }\right)
=1-$ $\alpha $, where $\hat{m}\left( x\right) $ is an estimator of $m\left(
x\right) $ and $\alpha \in \left( 0,1\right) $ is a pre-specified error
probability.

One classical approach to construct simultaneous confidence intervals is
to first obtain the asymptotic distribution of $\left[ \hat{m}\left(
x\right)\!-\!m\left( x\right) \right]\!/\!\sqrt{\limfunc{var}\{\hat{m}\left(
x\right) \}}$ which is often the standard normal distribution so that the
pointwise confidence intervals for $m\left( x\right) $ are constructed. Then
one can establish simultaneous confidence intervals for the values of the
regression curve at the design points by Bonferroni's Inequality. One
serious drawback of this approach is that the simultaneous confidence
intervals are too conservative. Johnston (1982) and H\"{a}rdle (1989) made a
substantial improvement through studying the limiting distribution of the
maximal deviation $\sup_{x\in \left[ a,b\right] }\left\vert \hat{m}\left(
x\right) -m\left( x\right) \right\vert $ for the kernel estimator and later
Wang and Yang (2009) extended the results to the B spline regression. As
formulated in the above works, Zheng et al. (2014) derived an SCB for the
mean function of sparse functional data and Gu et al. (2014) considered an
SCB for varying coefficient regression with sparse functional data, and
Zheng et al. (2016) studied an SCB for generalized additive models.
Furthermore, Gu and Yang (2015) proposed an SCB for the single-index link
function, and Song and Yang (2009), Cai and Yang (2015) and Cai et al.
(2019b) studied SCBs for the variance function $\sigma ^{2}\left( x\right) $%
. In addition, H\"{a}rdle and Marron (1991) and Claeskens and Van Keilegom
(2003) proposed bootstrap SCBs for $m\left( x\right) $ based on kernel
regression, and Eubank and Speckman (1993), Hall and Titterington (1988),
Wang (2012), Cai et al. (2014), and Cai et al. (2019b) investigated SCBs for
$m\left( x\right) $ in nonparametric regression with equally spaced design.

All the above and other related works on SCBs for nonparametric regression
are for fully observed data. To the best of our knowledge, there are no
related works on SCBs for the data with partially missing observations which
is a common situation in applications; see Little and Rubin (2019) for an
introduction on missing data and many examples. When the data are not
missing completely at random, using the complete case analysis by simply
discarding the missing data can lose efficiency and yield inconsistent
estimates since the conditional distribution of the response given the
observed covariates is in general not equal to the underlying true
conditional distribution of the response given all the covariates.

A series of efforts have been made to deal with missing data. The main
approaches include likelihood method, inverse selection probability weighted
approach, imputation and EM algorithm. For example, Qin et al. (2009)
considered likelihood approach, while Wang et al. (1997, 1998), Lipsitz et
al. (1999) and Liang et al. (2004) studied an inverse selection probability
weighting method. Hsu et al. (2014) proposed a nearest neighbor-based
nonparametric multiple imputation approach to recover missing covariate
information. Chen and Little (1999) applied the EM algorithm. See also
Ibrahim et al. (2005), Kim and Shao (2013) and Little and Rubin (2019) for
comprehensive overviews of statistical methods handling missing data.
However, most of these existing works mainly study the consistency and
asymptotic properties at any fixed point of the proposed estimator.

In this paper, we study the global inference for the mean function $m\left(
x\right) $ by constructing an asymptotically accurate SCB when covariates
are missing at random (MAR) meaning that the missingness mechanism depends
only on variables that are fully observable. We employ a weighted estimator
for $m\left( x\right) $ based on the inverse selection probability weights,
which is shown to be oracally efficient in the sense that the estimator with
estimated selection probabilities under a correctly specified model is
uniformly as efficient as that with true selection probabilities. The
asymptotic distribution of the maximal deviation of the estimator from the
true mean function is provided and hence an asymptotically accurate SCB for $%
m(x)$ is constructed.

As an illustration, our proposed SCB is applied to the data collected from
the Canada 2010/2011 Youth Student Survey to study the relationship between
self-esteem and BMI. Figure \ref{Fig:plot_SCB_real_data} depicts the
weighted local linear estimator and the SCB for the data. The null
hypothesis of the mean function $m(x)=a+bx$ for some constants $a$ and $b$
is tested by our SCB with the minimum confidence level covering the null
curve being $67.7\%$. Hence, with the $p$-value of $0.323$ one cannot reject
the null hypothesis; see Section 6 for more details.

The rest of the paper is organized as follows. Section 2 presents the main
theoretical results and the detailed procedure to implement the proposed method. Finite
sample simulation results and real data analyses are reported in Sections 3
and 4, respectively. Proofs of the main results are provided in the Appendix and the online Supplementary Materials.

\fontsize{12}{14pt plus.8pt minus .6pt}\selectfont
\setcounter{chapter}{2} 
\vskip 2mm \noindent {\large \textbf{2. Main Results\label{SEC:Main results}}
} \vskip 2mm \noindent {\large \textbf{2.1 A new SCB for the mean function} }

When samples $\left( X_{i},Y_{i}\right) $ are fully observable, Fan and
Gijbels (1996) proposed the local linear regression method to estimate $%
m\left( x\right) $ by solving
\begin{equation}
\text{argmin}_{\beta _{0},\beta _{1}\in \mathbb{R}}n^{-1}\sum%
\limits_{i=1}^{n}\left\{ Y_{i}-\beta _{0}-\beta _{1}\left( X_{i}-x\right)
\right\} ^{2}K_{h}\left( X_{i}-x\right) ,  \label{EQ:loclinear}
\end{equation}%
where 
$K_{h}\left( \cdot\right) =h^{-1}K\left( \cdot/h\right) $ is a rescaled
kernel function 
with bandwidth $h$. However, when covariates are MAR, the complete case
analysis in (\ref{EQ:loclinear}) by using only fully observed ($X_{i}$, $%
Y_{i}$) can result in biased estimator for $m(x)$. Assume that the observed
data are $(\delta_i, \delta_i X_i, Y_i)$, $i=1, \dots, n$, where $\delta
_{i}=1$ if $X_{i}$ is observed and $\delta _{i}=0$ otherwise, and $\pi
_{i}=P\left( \delta _{i}=1|Y_{i},X_{i}\right) =P\left( \delta
_{i}=1|Y_{i}\right) =\pi \left( Y_{i}\right) $ is the selection probability
by our MAR assumption. To accommodate the missingness, we apply the
Horvitz-Thompson inverse selection weighted method by minimizing the
following quantity with respect to $(\beta _{0},\beta _{1}),$
\begin{equation}
n^{-1}\sum\limits_{i=1}^{n}\frac{\delta _{i}}{\pi _{i}}\left\{ Y_{i}-\beta
_{0}-\beta _{1}\left( X_{i}-x\right) \right\} ^{2}K_{h}\left( X_{i}-x\right)
,  \label{EQ:weight_loclinear}
\end{equation}%
By least squares, one obtains the estimator $\hat{m}\left( x,\pi \right) $
for $m(x)$ with
\begin{equation}
\hat{m}\left( x,\pi \right) =e_{0}^{T}\left( \mathbf{X}^{T}\mathbf{WX}%
\right) ^{-1}\mathbf{X}^{T}\mathbf{WY,}  \label{EQ:represen_mahtpi}
\end{equation}%
where
\begin{equation*}
\mathbf{X=}\left(
\begin{array}{ccc}
1 & \cdots & 1 \\
X_{1}-x & \cdots & X_{n}-x%
\end{array}%
\right) ^{T},
\end{equation*}%
$\mathbf{W=}\frac{1}{n}\limfunc{diag}\left( \frac{\delta _{1}}{\pi _{1}}%
K_{h}\left( X_{1}-x\right) ,\dots ,\frac{\delta _{n}}{\pi _{n}}K_{h}\left(
X_{n}-x\right) \right) $, $\mathbf{Y}=\left( Y_{1},\dots ,Y_{n}\right) ^{T}$%
, and $e_{0}=\left( 1,0\right) ^{T}$. Here $\hat{m}\left( x,\pi \right) $ is
used to emphasize its dependence on the selection probability function $\pi
(y)$.

Note that the selection probability function $\pi (y)$ is generally unknown.
Here we assume that $\pi \left( y\right) $ follows a parametric binary model
$\pi \left( y,\mathbf{\alpha }\right) $ where $\mathbf{\alpha }$ is some
unknown parameter vector. For example, assuming a logistic regression model,
$\pi_{i}=\pi \left( Y_{i},\mathbf{\alpha }\right) =P(\delta
_{i}=1|Y_{i})=\{1+\exp \left( -\alpha _{0}-\alpha _{1}Y_{i}\right) \}^{-1}$,
$\mathbf{\alpha =}\left( \alpha _{0},\alpha _{1}\right) ^{T}$. By applying
the maximum likelihood approach, one easily obtains a root-$n$ consistent
estimate $\mathbf{\hat{\alpha}}$; see Robins et al. (1994) and Wang et al.
(1998) for related studies and Hosmer and Lemeshow (2005) for a global
statistic test for examining the pre-assumed binary regression model. Denote
the resulting selection probability function estimator as $\hat{\pi}\left(
y\right) =\hat{\pi}(y,\mathbf{\hat{\alpha}})$ and let $\hat{\pi}_{i}=\hat{\pi%
}\left( Y_{i},\mathbf{\hat{\alpha}}\right) $, $i=1, \dots, n$. Thus,
replacing $\pi _{i}$ in (\ref{EQ:weight_loclinear}) with $\hat{\pi}_{i}$,
the feasible weighted estimator $\hat{m}\left( x,\hat{\pi}\right) $ of $%
m\left( x\right) $ is derived with
\begin{equation}
\hat{m}\left( x,\hat{\pi}\right) =e_{0}^{T}\left( \mathbf{X}^{T}\mathbf{\hat{%
W}X}\right) ^{-1}\mathbf{X}^{T}\mathbf{\hat{W}Y,}
\label{EQ:represen_mahtpihat}
\end{equation}%
where the symbols with a hat on the right side of the equation above are the
same as those in equation (\ref{EQ:represen_mahtpi}) but with $\pi _{i}$
replaced by $\hat{\pi}_{i}$.

For any function $\phi \left( x\right) $, we use $\phi ^{\left( s\right)
}(x) $ to represent its $s$-th order derivative, and for any integer $p\geq
0 $ and use $C^{\left( p\right) }\left[ c,d\right] $ to indicate the space
of functions that have continuous $p$-th derivative on the interval $\left[
c,d\right] $ with letting $C\left[ c,d\right] =C^{\left( 0\right) }\left[ c,d%
\right] $. For any real positive sequences $l_{n}$ and $d_{n},$ $l_{n}\ll
d_{n}$ means $l_{n}/d_{n}\rightarrow 0$ as $n\rightarrow \infty $.

To construct an accurate SCB for the mean function $m(x)$, we need the
following general assumptions:

\begin{enumerate}
\item[(A1)] \textit{The mean function }$m\left( x\right) \in C^{\left(
2\right) }\left[ a,b\right] $\textit{\ and the density function }$%
f_{X}\left( x\right) $ \textit{of }$X$\textit{\ is positive} \textit{in the
open interval }$\left( a,b\right) $\textit{\ with} $f_{X}\left( x\right) \in
C^{(1)}\left[ a,b\right] $.\textit{\ Moreover, the joint density function $%
f_{X,\varepsilon }\left( x,\varepsilon \right) $ \textit{of} $\left(
X,\varepsilon \right) $} \textit{\ has continuous first order partial
derivative with respect to }$x$.

\item[(A2)] \textit{The variance function $\sigma ^{2}(x)$ is bounded on $%
\left[ a,b\right] $ and\ }$\tint \varepsilon ^{2}f_{X,\varepsilon |\delta
=1}\left( x,\varepsilon \right) d\varepsilon $ \textit{\ has a positive
lower bound for all }$x\in \left[ a,b\right] $, \textit{where} $f_{X,\varepsilon
|\delta =1}\left( x,\varepsilon \right) $ \textit{is the joint density
function of $\left( X,\varepsilon \right) $ given $\delta =1$. In addition,
there exist constants $\eta >4$\textit{\ and }$M_{\eta }>0$ \textit{such that%
} $\func{E}(\left\vert \varepsilon \right\vert ^{2+\eta }\big\vert X)\leq M_{\eta }$
a.s.}

\item[(A3)] \textit{The kernel function }$K\left( \cdot \right) $ \textit{is
a symmetric probability density function supported on }$\left[ -1,1\right] $
and $\in C^{\left( 1\right) }\left[ \mathbb{-}1,1\right] $\textit{.}

\item[(A4)]
\textit{The selection probability function $\pi \left( y\right) $ follows a
parametric binary model and has} \textit{a positive lower bound }$c_{\pi }$%
\textit{. Moreover, it has bounded second order partial derivative with
respect to }$y$\textit{\ and has bounded first order partial derivative with
respect to }$\mathbf{\alpha .}$

\item[(A5)] \textit{The bandwidth }$h=h_{n}$ \textit{satisfies }$%
n^{-1/3}\log n\ll h\ll n^{-1/5}\log ^{-1/5}n$.
\end{enumerate}

Assumptions (A1)--(A3) are elementary conditions in nonparametric kernel
regression adapted from Johnston (1982), H\"{a}rdle (1989), Wang and Yang
(2009), and Cai et al. (2019b). Assumption (A1) implies that for any compact
subinterval $\left[ a_{0},b_{0}\right] \subset \left( a,b\right) $, there
exist positive constants $c_{f},C_{f}$ such that $c_{f}\leq f_{X}\left(
x\right) \leq C_{f}$, $\forall x\in \left[ a_{0},b_{0}\right] $. The
condition $\eta >4$ in Assumption (A2) can be relaxed to $\eta >3$ but then
the lower order restriction of the bandwidth is more complicated. For
simplicity here we use $\eta >4$. Assumption (A3) entails that $\mu
_{0}\left( K\right) =1$ and $\mu _{1}\left( K\right) =0$ where $\mu
_{l}\left( K\right) =\tint\nolimits_{-1}^{1}u^{l}K\left( u\right) du,l=0,1$.
Assumption (A4) is typical in the missing data analysis. The same condition
appears in Liang et al. (2004) and Wang et al. (1997). Assumption (A5) is
about the choice of bandwidth $h$. Technically, it keeps the bias at a lower
rate than the variance and entails some negligible nonlinear remainder terms.

For any functions $\varphi _{n}\left( x\right) $ and $\phi _{n}\left(
x\right) ,x\in \mathcal{D}$, we use $\varphi _{n}\left( x\right) =$ $O\left(
\phi _{n}\left( x\right) \right) $ and $\varphi _{n}\left( x\right) =o\left(
\phi _{n}\left( x\right) \right) $ to mean \textquotedblleft $\varphi
_{n}\left( x\right) /\phi _{n}\left( x\right) $ is bounded and $\varphi
_{n}\left( x\right) /\phi _{n}\left( x\right) $ tends to $0$ as $%
n\rightarrow \infty $ for any fixed $x\in \mathcal{D}$", while use $\varphi
_{n}\left( x\right) =$ $U\left( \phi _{n}\left( x\right) \right) $ and $%
\varphi _{n}\left( x\right) =u\left( \phi _{n}\left( x\right) \right) $ to
mean \textquotedblleft $\varphi _{n}\left( x\right) /\phi _{n}\left(
x\right) $ is bounded and $\varphi _{n}\left( x\right) /\phi _{n}\left(
x\right) $ tends to $0$ as $n\rightarrow \infty $ for all $x\in \mathcal{D}$
uniformly". We use $O_{p}$, $o_{p}$, $U_{p}$ and $u_{p}$ to represent the
corresponding terms in probability.

\begin{theorem}
\label{THM:mhatpi_m} Under Assumptions (A1)--(A5), as $n\rightarrow \infty $%
, for any $x\in \left[ a_{0},b_{0}\right] $, one has
\begin{equation*}
\hat{m}\left( x,\pi \right) -m\left( x\right) =R_{n}\left( x\right)
+2^{-1}h^{2}\mu _{2}\left( K\right) m^{\left( 2\right) }\left( x\right)
\!+u_{p}\left( h^{2}\right),
\end{equation*}%
where $R_{n}\left( x\right) =n^{-1}f_{X}^{-1}\left( x\right)
\sum\limits_{i=1}^{n}\frac{\delta _{i}}{\pi _{i}}K_{h}\left(
X_{i}\!-\!x\right) \varepsilon _{i}$.
\end{theorem}

The proof of Theorem \ref{THM:mhatpi_m} is given in the Appendix. Together
with $h\ll n^{-1/5}\log ^{-1/5}\!n$, one can easily obtain that
\begin{equation}
\sup_{x\in \left[ a_{0},b_{0}\right] }\left\vert \hat{m}\left( x,\pi \right)
-m\left( x\right) \right\vert =\sup_{x\in \left[ a_{0},b_{0}\right]
}\left\vert R_{n}\left( x\right) \right\vert +O_{p}\left(
h^{2}\right) .  \label{EQ:supmtilde-m}
\end{equation}%
Lemma S.2 
in the Supplementary Materials shows that $\sup_{x\in \lbrack
a_{0},b_{0}]}\!\left\vert R_{n}(x)\right\vert\!=\!O_{p}(n^{-1/2}h^{-1/2}\!\log
^{1/2}\!n)$ which implies that the dominating term of $\sup_{x\in \lbrack
a_{0},b_{0}]}$ $\left\vert \hat{m}\left( x,\pi \right) -m\left( x\right)
\right\vert $ is $\sup_{x\in \lbrack a_{0},b_{0}]}\left\vert
R_{n}(x)\right\vert $.


Let $\Delta _{n}=\sum_{i=1}^{n}\delta _{i}$ 
be the number of complete cases and denote the ratio by $r_{n}=\Delta _{n}/n$%
. Since $\delta _{1},\dots,\delta _{n}$ are i.i.d., it is readily seen that
\begin{equation}
r_{n}=P\left( \delta _{1}=1\right) +O_{p}(n^{-1/2}).  \label{limit_for_rn}
\end{equation}
We now give the following theorem which describes the limiting distribution
of the maximal deviation between $\hat{m}\left( x,\pi \right) $ and $m\left(
x\right)$. Its proof is given in the Appendix.

\begin{theorem}
\label{THM:mhatpi_m_distribution} Under Assumptions (A1)--(A5), as $%
n\rightarrow \infty $, for any $t\in \mathbb{R},$%
\begin{gather}
P\left\{ a_{h}\left[ \sup_{x\in \left[ a_{0},b_{0}\right] }\left\vert \frac{%
\left( nh\right) ^{1/2}r_{n}^{-1/2}\left\{ \hat{m}\left( x,\pi \right)
-m\left( x\right) \right\} }{d^{1/2}\left( x\right) }\right\vert -b_{h}%
\right] \leq t\right\}  \notag \\
\rightarrow \exp \left\{ -2\exp \left( -t\right)
\right\} ,  \label{EQ:limit_dist}
\end{gather}%
where%
\begin{gather*}
a_{h}=\sqrt{-2\log \left( h/\left( b_{0}-a_{0}\right) \right) }%
,b_{h}=a_{h}+2^{-1}a_{h}^{-1}\log \left( 4^{-1}\pi ^{-2}C\left( K\right)
\right) , \\
d\left( x\right) =\lambda \left( K\right) s\left( x\right) f_{X}^{-2}\left(
x\right) ,s\left( x\right) =\int \frac{\varepsilon ^{2}}{\pi ^{2}\left(
m\left( x\right) +\varepsilon \right) }f_{X,\varepsilon |\delta =1}\left(
x,\varepsilon \right) d\varepsilon , \\
\lambda \left( K\right) =\tint\nolimits_{-1}^{1}K^{2}\left( u\right)
du,C\left( K\right) =\lambda ^{-1}\left( K\right)
\tint\nolimits_{-1}^{1}\left\{ K^{\left( 1\right) }\left( u\right) \right\}
^{2}du.
\end{gather*}
\end{theorem}

Note that when data are fully observed, i.e., $\pi \left( y\right) \equiv
1,r_n\equiv 1$, $s\left( x\right) $ becomes $\sigma ^{2}\left( x\right)
f_{X}\left( x\right) $. In such a case, the result degenerates to that for
the local linear estimator for fully observed data, which extends the result
of Johnston (1982) for the Nadaraya-Watson kernel estimator to the local
linear estimator under more general conditions.

The proof of Theorem \ref{THM:mhatpi_m_distribution} is quite involved. It
uses the total probability formula that the probability of the dominating
term $\sup_{x\in \left[ a_{0},b_{0}\right] }\!\left\vert \left( nh\right)
^{1/2}r_{n}^{-1/2}\right.$ $\left. R_{n}(x) /d^{1/2}\left( x\right)
\right\vert$ is the weighted average of its conditional probability given $%
\Delta _{n}=n_0$ with weights $P(\Delta _{n}=n_0),n_0=0,1,2,...,n$; see the
detailed argument in the Appendix. In the rest theoretical development, we
assume that the parametric model for $\pi $ is correctly specified so that
the estimator
$\mathbf{\hat{\alpha}}$
satisfies $\mathbf{\hat{\alpha}}-\mathbf{\alpha }=O_{p}(n^{-1/2})$. Theorem %
\ref{THM:mhatpi_mhatpihat_oracle} below compares the difference between the
estimator based on the true selection probability function $\pi $ and that
based on the estimated selection probability function $\hat{\pi}$. Its proof
is given in the Appendix.

\begin{theorem}
\label{THM:mhatpi_mhatpihat_oracle} Under Assumptions (A1)--(A5), as $%
n\rightarrow \infty $,
\begin{equation*}
\sup_{x\in \left[ a_0,b_0\right] }\left\vert \hat{m}\left( x,\hat{\pi}%
\right) -\hat{m}\left( x,\pi \right) \right\vert =O_{p}\left(
n^{-1/2}\right) .
\end{equation*}
\end{theorem}

Combining Theorems \ref{THM:mhatpi_m_distribution} and \ref%
{THM:mhatpi_mhatpihat_oracle} and Slutsky's Theorem, one obtains the
following result:

\begin{theorem}
\label{THM:SCB_feasible} Under Assumptions (A1)--(A5), as $n\rightarrow
\infty $, for any $t\in \mathbb{R},$%
\begin{gather*}
P\left\{ a_{h}\left[ \sup_{x\in \left[ a_{0},b_{0}\right] }\left\vert \frac{%
\left( nh\right) ^{1/2}r_{n}^{-1/2}\left\{ \hat{m}\left( x,\hat{\pi}\right)
-m\left( x\right) \right\} }{d^{1/2}\left( x\right) }\right\vert -b_{h}%
\right] \leq t\right\}\\
 \rightarrow \exp \left\{ -2\exp \left( -t\right)
\right\} .
\end{gather*}
\end{theorem}

Theorem \ref{THM:SCB_feasible} above can be used to construct a theoretical
SCB for $m\left( x\right) $ which depends on unknown quantity $d\left(
x\right) $. To obtain a feasible SCB, we estimate $d\left( x\right) $ by
\begin{equation*}
\hat{d}_{n}\left( x\right) =\Delta _{n}^{-1}h\hat{f}_{X}^{-2}\left( x\right)
\dsum\limits_{i=1}^{n}\frac{\delta _{i}}{\hat{\pi}_{i}^{2}}K_{h}^{2}\left(
X_{i}\!-\!x\right) \hat{\varepsilon}_{i}^{2},
\end{equation*}%
where $\hat{\varepsilon}_{i}=Y_{i}-\hat{m}\left( X_{i},\hat{\pi}_{i}\right) $
and $\hat{f}_{X}\left( x\right) $ is the weighted kernel density pilot
estimator of $f_{X}\left( x\right) $ with
\begin{equation}
\hat{f}_{X}\left( x\right) =n^{-1}\sum\limits_{i=1}^{n}\frac{\delta _{i}}{%
\hat{\pi}_{i}}K_{h_{f}}\left( X_{i}\!-\!x\right) ,  \label{EQ:density_est}
\end{equation}%
in which we recommend to use the Silverman's rule-of-thumb bandwidth
(Silverman (1986), p.48) computed with complete data for $h_{f}$ which has the order of $n^{-1/5}$.

\begin{theorem}
\label{THM:duhat_du} Under Assumptions (A1)--(A5), as $n\rightarrow \infty $%
, one has%
\begin{equation*}
\sup_{x\in \left[ a_{0},b_{0}\right] }\left\vert \hat{d}_{n}\left( x\right)
-d\left( x\right) \right\vert =O_{p}\left( n^{-1/2}h^{-3/2}\log^{1/2}n\right) \text{.}
\end{equation*}
\end{theorem}
Note that $n^{-1/2}h^{-3/2}\log^{1/2}n\ll \log^{-1} n$ by Assumption (A5). Therefore, we have the following corollary.

\begin{corollary}
\label{COL:SCB} Under Assumptions (A1)--(A5), for any $\alpha \in \left(
0,1\right) $, an asymptotic $100\left( 1-\alpha \right) \%$ simultaneous
confidence band for $m\left( x\right) $ over $\left[ a_{0},b_{0}\right] $ is
\begin{equation}
\hat{m}\left( x,\hat{\pi}\right) \pm \left( nh\right) ^{-1/2}r_{n}^{1/2}\hat{%
d}_{n}^{1/2}\left( x\right) \left( b_{h}+a_{h}^{-1}q_{\alpha }\right) ,
\label{EQ:feasible_SCB}
\end{equation}%
where $q_{\alpha }=-\log \left\{ -\frac{1}{2}\log \left( 1-\alpha \right)
\right\} $ and $a_{h}$, $b_{h}$ are given in Theorem \ref%
{THM:mhatpi_m_distribution}.
\end{corollary}

\vskip 2mm \noindent {\large \textbf{2.2 Implementation} }

In this subsection, we describe the detailed procedure to implement the
asymptotic SCB in (\ref{EQ:feasible_SCB}). They will be used throughout
Section 3 and Section 4 for simulation studies and real data analysis.

The range of the covariate variable is taken as ${\small [\hat{a},\hat{b}]}$
with $\hat{a}=\min_{\{1\leq i\leq n,\,\delta _{i}=1\}}$ $X_{i}$ and $\hat{b}%
=\max_{\{1\leq i\leq n,\,\delta _{i}=1\}} X_i $, while the compact
subinterval ${\small [\hat{a}_{0},\hat{b}_{0}]}$ with $\hat{a}_{0}=0.9\hat{a}%
+0.1\hat{b}$ and $\hat{b}_{0}=0.9\hat{b}+0.1\hat{a}$ is regarded as the
interval over which the SCBs are constructed. The quartic kernel, $K\left(
u\right) =15\left( 1-u^{2}\right) ^{2}I\left( \left\vert u\right\vert \leq
1\right)$ $ /16$, is used for the weighted local linear estimator in (\ref%
{EQ:represen_mahtpihat}) and the weighted kernel density estimator in (\ref%
{EQ:density_est}), satisfying Assumption (A3).

Regarding the bandwidth selection for $\hat m$ in (\ref{EQ:represen_mahtpi}%
), we adopt $h=h_{\text{rot}}\log^{-\rho}n$ for some $\rho>1/5$, where $h_{%
\text{rot}}$ is the rule-of-thumb bandwidth in Fan and Gijbels (1996,
Equation (4.3)) computed with the complete data. Note that the order of $h_{%
\text{rot}}$ is $n^{-1/5}$ and hence the order of $h$ is $%
n^{-1/5}\log^{-\rho}n$ which satisfies Assumption (A5). We have found in
extensive simulations that $h = h_{\text{rot}} \log^{-1/4}n$ (i.e., $\rho =
1/4$) works quite well and that is what we recommend.

\vskip 2mm \noindent {\large \textbf{3. Simulation Studies \label%
{SEC:simulation}} }

In this section, we investigate the finite sample behaviors of the proposed
SCB and the finite sample effect due to estimating the selection
probabilities. For comparison, we also list the results of the complete case
SCB for local linear regression by directly ignoring the missing covariates,
denoted by SCB-CC.

The following four cases were examined:
\begin{eqnarray*}
\text{Case 1} &\text{:}&m\left( X\right) =\sin \left( \pi X\right) ,\sigma
\left( X\right) =1; \\
\text{Case 2} &\text{:}&m\left( X\right) =\sin \left( \pi X\right) ,\sigma
\left( X\right) =2\exp (X)\left\{ \exp (X)+1\right\} ^{-1}; \\
\text{Case 3} &\text{:}&m\left( X\right) =\exp (-6X^{3}/5),\sigma \left(
X\right) =1; \\
\text{Case 4} &\text{:}&m\left( X\right) =\exp (-6X^{3}/5),\sigma \left(
X\right) =2\exp (X)\left\{ \exp (X)+1\right\} ^{-1},
\end{eqnarray*}%
where $X\sim U\left[ -1,1\right]$, and the error $\varepsilon \sim N\left(
0,\sigma ^{2}\left( x\right) \right) $. Clearly, these scenarios include
both homoscedastic errors (Case 1, Case 3) and heteroscedastic errors (Case
2, Case 4). 
Two models for the selection probability function were considered: (i)
logistic model $\pi\left(Y\right)=P\left(\delta =1|Y\right) =\left\{ 1+\exp(
-\alpha _{0}-\alpha _{1}Y) \right\}^{-1}$, and (ii) probit model $%
\pi\left(Y\right)=P\left( \delta =1|Y\right) =\Phi \left(\alpha _{0}^{\ast
}+\alpha _{1}^{\ast }Y\right)$, where $\Phi $ is the standard normal
cumulative distribution function. We took $\left( \alpha _{0},\alpha
_{1}\right)$ and $\left( \alpha _{0}^{\ast },\alpha _{1}^{\ast }\right)$ as $%
(1.8,1)$ and $(1,0.5)$, respectively, leading to approximately $8\%$ to $20\%$ of the
data missing (low proportion of missing). We also took $\left( \alpha
_{0},\alpha _{1}\right)$ and $\left( \alpha _{0}^{\ast },\alpha _{1}^{\ast
}\right)$ as $(0.2,0.6)$ and $(0.1,0.3)$, respectively, leading to
approximately $31\%$ to $46\%$ of the data missing (high proportion of missing). The
sample sizes were $n=400,600,800$ and the confidence levels were $1-\alpha
=0.95,0.99$.

We first look at the performance of the proposed SCB in the cases where the
selection probability models are correctly specified. Tables \ref%
{TAB:coverage_logit_2_1}--\ref{TAB:coverage_probit_0502} give the coverage
frequencies over 1000 replications that the true mean function was covered
by the SCB in (\ref{EQ:feasible_SCB}) and the SCB-CC at the equally spaced
points $\hat{a}_{0}+(\hat{b}_{0}-\hat{a}_{0})k/400$, $k=0,\dots,400$. One
can see that for all scenarios, the coverage frequencies of the proposed SCB
in (\ref{EQ:feasible_SCB}) are close to the nominal confidence levels $0.95$
and $0.99$ while the coverage frequencies of SCB-CC are far lower than the
nominal levels, and the average lengths of SCB-CC are systematically
narrower than that of the proposed SCB. Meanwhile, the average lengths of
the SCBs decrease as the sample size $n$ increases, as expected. 
All in all, it can be seen that the proposed SCB in (\ref{EQ:feasible_SCB})
performs much better than the SCB-CC. This is because the local linear
estimation in the complete case is generally biased for the underlying true
function. These findings confirm our theoretical results.

We next investigate the sensitivity of the SCB to the selection probability
model misspecification. Similar to Wang et al. (1997), we carried out a
simulation study which is similar to that in Table \ref%
{TAB:coverage_logit_0206} except that the selection probability is truncated
above by $0.75$. As a result, about $46\%$ of the cases had missing
covariates. Therefore, using the logistic regression model to fit $\pi(y)$
is not completely correct in this setting. Table \ref%
{TAB:coverage_misprobit_05_02} summarizes the simulation results under this
misspecification. One can see that for all the scenarios, the coverage
frequencies are quite close to those under the correct specification of $\pi
(y)$. It suggests that the proposed SCB is not very sensitive to
misspecification of the selection probability function.

To visualize the SCB for the mean function, Figures \ref%
{Fig:plot_SCB_logit0206_case2} and \ref{Fig:plot_SCB_logit0206_case4} were
created based on two samples of size 400 and 800 for Case 1 and Case 4 under
the logit missing mechanism with $\left( \alpha _{0},\alpha
_{1}\right)=(0.2,0.6)$. 
One can see that the SCB for $n=800$ is narrower and fits the true mean
function better than those for $n=400$, which corroborates our
asymptotically theoretical results. We have also created the figures in
other cases, and the results are similar.

\vskip 2mm \noindent {\large \textbf{4. Real Data Analysis \label%
{SEC:real_data_analysis}}}

In this section, we illustrate an application to the data from the Canada
2010/2011 Youth Student Survey. The 2010/2011 Youth Student Survey sponsored
by Health Canada is a pan-Canadian, classroom-based survey on a
representative youth students in grades $6$--$12$ between October 2010 and
June 2011. It aims to provide Health Canada, provinces, schools,
communities, and parents with timely and reliable data on tobacco, alcohol
and drug use in addition to other related issues about Canadian students;
see more details in \textit{2010-2011 YSS Student Survey Data Codebook} or
from \textit{%
https://uwaterloo.ca/canadian-student-tobacco-alcohol-drugs-survey}.

We focused on a subset of the data collected from white female youth
students in grades $6$--$12$ to study the relationship between self-esteem
and Body Mass Index (BMI); see the interesting related discussions and
further references in Habib et al. (2015) and Al Ahmari et al. (2017). In
this data set, the self-esteem was measured by using a score ranging from $0$
to $12$, and the BMI was computed by the weight over height in meter
squared, ranging from $10.04$ to $49.78$. There were a total of $5343$
students with having complete observations on self-esteem, while only $3565$
students provided BMI ($33.2\%$ missing rate).

For the data missingness mechanism, we used the logistic regression to
estimate $\pi(y)$. The fitted estimates are $\hat{\mathbf{\alpha}}=(
0.82585, -0.015)^{T} $. To further judge how well the model fits the
missingness pattern in the data, the Hosmer-Lemeshow goodness of fit test
was employed with the $p$-value $=0.17$. Thus one cannot reject the null
hypothesis that the logistic model is correct. Figure \ref%
{Fig:plot_SCB_real_data} shows the inverse selection probability weighted
local linear estimate $\hat{m}(x,\hat{\pi})$ (thick solid line) and the $%
95\% $ and $65.6\%$ SCBs (solid lines). The SCB was applied to test the null
hypothesis $H_0:m(x)=a+bx$ where the coefficients $(a,b)^T$ were computed by
the inverse selection probability weighted least square method; see the null
curve (dashed line) in Figure \ref{Fig:plot_SCB_real_data}. One can see that
the null curve is completely covered by the $95\%$ SCB. Thus the null
hypothesis of the mean function being a linear function cannot be rejected
at the significant level $=0.05$. Applying Theorem \ref{THM:SCB_feasible},
we obtained that the minimum confidence level containing the null curve is $%
67.7\%$; see the right panel of Figure \ref{Fig:plot_SCB_real_data}.
Therefore, the null hypothesis cannot be rejected with $p$-value $=0.323$.

Moreover, one can also see that the mean curve has a general decreasing
trend, i.e., there is a negative association between self-esteem and BMI
among white female youth students in grades $6$--$12$. This result agrees
with that discovered by Habib et al. (2015). Meanwhile, according to Habib
et al. (2015), a BMI between $20$ and $25$ is considered normal, a BMI
between $25$ and $30$ is considered overweight, and a BMI $>30$ is
considered obese. Therefore, even if female students were within normal
weight range, their self-esteem was still decreased as BMI increased. This
may be because female students in general are more likely to see themselves
as obese or overweight and show dissatisfaction with their body image even
if they have a healthy weight.

\vskip 2mm \noindent {\large \textbf{5. Concluding Remarks}}

In this paper, an asymptotically accurate SCBs were constructed for the
nonparametric mean function with covariates missing at random by employing
the weighted estimator based on inverse selection probabilities. The
limiting distribution of the global estimation error (also known as maximal
deviation) was derived, overcoming the main technical challenge on
formulating such a confidence band. The proposed estimator for the mean
function was shown to be oracally efficient in the sense that using root-$n$
consistent selection probability estimates is as efficient as that when the
selection probabilities were known as a prior. Simulation studies confirm
our theoretical findings and the analysis of the Canada 2010/2011 Youth
Student Survey data illustrates the versatility of the SCB. The methodology
should also be suitable to partial linear models for missing covariates data
(Wang (2009)). Further investigations may lead to similar constructions of
SCBs for generalized nonparametric models, partial linear single-index
models and functional data with covariates missing at random.

\vskip 2mm \noindent {\large \textbf{5. Supplementary Materials}}

The online Supplementary Materials contain some lemmas to be used
in the proofs of the main theorems in Section 2.

\vskip 2mm \noindent {\large \textbf{Acknowledgements}}

This research was supported in part by the National Natural Science
Foundation of China Award NSFC \#11701403, \#11901521, First Class Discipline of Zhejiang--A (Zhejiang Gongshang University--Statistics),  a grant from the Key Lab of Random Complex Structure and
Data Science, CAS, China, and the Simons Foundation
Mathematics and Physical Sciences Program Award \#499650.

\setcounter{section}{0}
\setcounter{equation}{0}
\setcounter{figure}{0}
\setcounter{theorem}{0}
\def\theequation{A.\arabic{equation}}
\def\thelemma{A.\arabic{lemma}}
\def\thetheorem{A.\arabic{theorem}}

\vskip 2mm \noindent {\large \textbf{Appendix}}

We use $a_{n}\sim b_{n}$ to represent $\lim_{n\rightarrow \infty
}a_{n}/b_{n}=c,$ where $c$ is some nonzero constant. For any function $%
\varphi \left( u\right) $ defined on $\left[ a,b\right] $, let $\left\Vert
\varphi \left( u\right) \right\Vert _{\infty }=\left\Vert \varphi
\right\Vert _{\infty }=\sup_{u\in \left[ a,b\right] }\left\vert \varphi
\left( u\right) \right\vert $.

\vskip 2mm \noindent {\large \textbf{A.1 Conditional limiting extreme value distribution of $R_{n}(x)$}}

This section contains the main steps to obtain the conditional extreme value distribution of $R_{n}(x)=n^{-1}f_{X}^{-1}\left( x\right)
\sum\limits_{i=1}^{n}\frac{\delta _{i}}{\pi _{i}}K_{h}\left(
X_{i}\!-\!x\right) \varepsilon _{i}$ shown in Theorem \ref{THM:Cond_R_n} at the end of this section which will be used in the total probability formula in the proof of Theorem \ref{THM:mhatpi_m_distribution}.

The Rosenblatt quantile transformation in Rosenblatt (1952) is
adopted with
\begin{equation*}
T\left( X,\varepsilon \right) =\left( X^{\ast },\varepsilon ^{\ast }\right)
=\left( F_{X|\delta =1}\left( X\right) ,F_{\varepsilon |X,\delta =1}\left(
\varepsilon |X\right) \right) ,
\end{equation*}%
where $F_{X|\delta =1}\left( X\right) $ is the conditional distribution
function of $X$ given $\delta =1$ and $F_{\varepsilon |X,\delta =1}\left(
\varepsilon |X\right) $ is the conditional distribution function of $%
\varepsilon $ given $X$ and $\delta =1$. This transformation produces
mutually independent uniform random variables $\left( X^{\ast },\varepsilon
^{\ast }\right) $ on $\left[ 0,1\right] ^{2}$. According to the strong
approximation theorem in Tusn\'{a}dy (1977) (Theorem 1), there exists a
sequence of two dimensional Brownian bridges $B_{n}$ such that
\begin{equation}
\sup\nolimits_{x,\varepsilon }\left\vert Z_{n}\left( x,\varepsilon \right)
-B_{n}\left( T\left( x,\varepsilon \right) \right) \right\vert =O\left(
n^{-1/2}\log ^{2}n\right) \text{ a.s.},  \label{EQ:strong_approximation}
\end{equation}%
where $Z_{n}\left( x,\varepsilon \right) =n^{1/2}\left\{ F_{n}\left(
x,\varepsilon \right) -F_{X,\varepsilon |\delta =1}\left( x,\varepsilon
\right) \right\} $ with $F_{n}\left( x,\varepsilon \right) $ and $%
F_{X,\varepsilon |\delta =1}$ $\left( x,\varepsilon \right) $ representing the
empirical and the theoretical distribution of $\left( X,\varepsilon \right) $
given $\delta =1$. The transformation and the strong approximation results
have been also used in Johnston (1982), H\"{a}rdle (1989), and Wang and Yang
(2009) for constructing SCBs for the nonparametric regression when data is
fully observed.

To obtain the distribution of $\sup_{x\in \left[ a_{0},b_{0}%
\right] }\left\vert R_n(x)\right\vert $ conditional on $\Delta _{n}=n_{0}$, we will show the following Lemmas \ref{LEM:process_aproxi}--\ref{LEM:xi1_xi2_approx_}. Here $\{ n_{0}\}$ is a sequence of numbers related to $n$ with $1\leq n_{0}\leq n$. By (7) it is clear that $n \to \infty $ if and only if $n_0 \to \infty$ in probability.  Meanwhile, due to the
i.i.d. assumption of the data, conditional on $\Delta _{n}=n_{0}$ is
equivalent to conditional on $\mathbf{\delta }_{n}=\left( \delta
_{1},...,\delta _{n}\right) ^{T}$ in which $n_{0}$ of the $\delta _{i}$
equal to 1 and the rest $\delta _{i}$ equal to 0. Without loss of
generality, let $\delta _{i}=1$ for $i=1,\dots ,n_{0}$ and $\delta _{i}=0$
for $i=n_{0}+1,\dots ,n$.

Notice that, for $i=1,\dots ,n$,
\begin{eqnarray*}
0 &=&\limfunc{E}\!\left\{ \frac{\delta _{i}}{\pi _{i}}K_{h}\left(
X_{i}\!-\!x\right) \varepsilon _{i}\right\} =\limfunc{E}\left[ \limfunc{E}%
\!\left\{ \left. \frac{\delta _{i}}{\pi _{i}}K_{h}\left( X_{i}\!-\!x\right)
\varepsilon _{i}\right\vert \delta _{i}\right\} \right] \\
&=&\limfunc{E}\left\{ \left. \frac{1}{\pi _{i}}K_{h}\left(
X_{i}\!-\!x\right) \varepsilon _{i}\right\vert \delta _{i}=1\right\} P\left(
\delta _{i}=1\right) .
\end{eqnarray*}%
Thus, conditional on $\Delta _{n}=n_{0}$, $1\leq n_{0}\leq n$, by symmetry
one has%
\begin{align*}
& \limfunc{E}\left\{ \left. \dsum\limits_{i=1}^{n}\frac{\delta _{i}}{\pi _{i}%
}K_{h}\left( X_{i}\!-\!x\right) \varepsilon _{i}\right\vert \Delta
_{n}=n_{0}\right\} \!\!\!\! \\
\!\!\!\!& =\limfunc{E}\left\{ \left. \dsum\limits_{i=1}^{n}\frac{\delta _{i}%
}{\pi _{i}}K_{h}\left( X_{i}\!-\!x\right) \varepsilon _{i}\right\vert \delta
_{1}=\cdots =\delta _{n_{0}}=1,\delta _{n_{0}+1}=\cdots =\delta
_{n}=0\right\} \\
\!\!\!\!& =n_{0}\limfunc{E}\left\{ \left. \frac{1}{\pi _{1}}K_{h}\left(
X_{1}\!-\!x\right) \varepsilon _{1}\right\vert \delta _{1}=1\right\} =0\text{
a.s.}
\end{align*}%
and
\begin{align}
& \limfunc{var}\!\left\{ \left. \tsum\limits_{i=1}^{n}\frac{\delta _{i}}{\pi
_{i}}K_{h}\left( X_{i}\!-\!x\right) \varepsilon _{i}\right\vert \Delta
_{n}=n_{0}\right\}  \notag \\
& =\limfunc{E}\left\{ \left. \{\dsum\limits_{i=1}^{n}\frac{\delta _{i}}{\pi
_{i}}K_{h}\left( X_{i}\!-\!x\right) \varepsilon _{i}\}^{2}\right\vert \delta
_{1}=\cdots =\delta _{n_{0}}=1,\delta _{n_{0}+1}=\cdots =\delta
_{n}=0\right\}  \notag \\
& =n_{0}\limfunc{E}\!\left( \left. \frac{1}{\pi _{1}^{2}}K_{h}^{2}\left(
X_{1}\!-\!x\right) \varepsilon _{1}^{2}\right\vert \delta _{1}=1\right)
\notag \\
& =n_{0}\int \frac{1}{\pi ^{2}\left( m\left( u\right) +\varepsilon \right) }%
K_{h}^{2}\left( u\!-\!x\right) \varepsilon ^{2}f_{X,\varepsilon |\delta
=1}\left( u,\varepsilon \right) dud\varepsilon  \notag \\
& =n_{0}h^{-1}\int \frac{1}{\pi ^{2}\left( m\left( x+hv\right) +\varepsilon
\right) }K^{2}\left( v\right) \varepsilon ^{2}f_{X,\varepsilon |\delta
=1}\left( x+hv,\varepsilon \right) dvd\varepsilon  \notag \\
& =n_{0}h^{-1}\int K^{2}\left( v\right) dv\int \frac{1}{\pi ^{2}\left(
m\left( x\right) +\varepsilon \right) }\varepsilon ^{2}f_{X,\varepsilon
|\delta =1}\left( x,\varepsilon \right) d\varepsilon \left\{ 1+u_{p}\left(
1\right) \right\}  \notag \\
& =n_{0}h^{-1}\lambda \left( K\right) s\left( x\right) \left\{ 1+u_{p}\left(
1\right) \right\} ,  \label{EQ:var_sumkepspi}
\end{align}%
where $\lambda \left( K\right) $ and $s\left( x\right) $ are given in Theorem 2. Note that conditional on $\Delta _{n}=n_{0}$,
without loss of generality, one can write
\begin{equation*}
R_{n}\left( x\right) =n^{-1}f_{X}^{-1}\left( x\right)
\sum\limits_{i=1}^{n_{0}}\frac{1}{\pi _{i}}K_{h}\left( X_{i}\!-\!x\right)
\varepsilon _{i}.
\end{equation*}

Conditional on $\Delta _{n}=n_{0}$ we now introduce the following
standardized stochastic process:
\begin{equation}
\zeta _{1n_{0}}\left( x\right) =\left( n_{0}h\right) ^{1/2}s^{-1/2}\left(
x\right) n_{0}^{-1}\tsum\limits_{i=1}^{n_{0}}\frac{1}{\pi _{i}}K_{h}\left(
X_{i}\!-\!x\right) \varepsilon _{i},  \label{DEF:zeta_1n0}
\end{equation}%
which can be rewritten as%
\begin{equation*}
\zeta _{1n_{0}}\left( x\right) =h^{1/2}s^{-1/2}\left( x\right) \int \int
\frac{1}{\pi \left( m\left( u\right) +\varepsilon \right) }K_{h}\left(
u-\!x\right) \varepsilon dZ_{n_{0}}\left( u,\varepsilon \right) ,
\end{equation*}%
where $Z_{n_{0}}\left( u,\varepsilon \right) $ is the same as $Z_{n}\left(
u,\varepsilon \right) $ in (\ref{EQ:strong_approximation}) but with $n$
replaced by $n_{0}$. %
%
%
%
%
%
%
%
%
%
%
%
%
%
%
%
%
%

Let $\kappa _{n}=n^{\theta }$ with $\frac{2}{3\eta }<\theta <\frac{1}{6}$
where $\eta >4$ is given in Assumption (A2), which together with Assumption
(A5) implies that
\begin{equation}
\kappa _{n}^{-\eta }h^{-2}\left( \log n\right) =O\left( 1\right) ,\text{ \ \
\ }\kappa _{n}^{2}n^{-1/2}h^{-1/2}\left( \log n\right) ^{5/2}=o\left(
1\right) .  \label{EQ:kn}
\end{equation}%
Then conditional on $\Delta _{n}=n_{0}$ one can define the following
processes to approximate $\zeta _{1n_{0}}\left( x\right) $:
\begin{eqnarray*}
\zeta _{2n_{0}}\left( x\right) &=&h^{1/2}s_{n}^{-1/2}\left( x\right) \int
\int_{\left\vert \varepsilon \right\vert \leq \kappa _{n}}\frac{1}{\pi
\left( m\left( u\right) +\varepsilon \right) }K_{h}\left( u-\!x\right)
\varepsilon dZ_{n_{0}}\left( u,\varepsilon \right) , \\
\zeta _{3n_{0}}\left( x\right) &=&h^{1/2}s_{n}^{-1/2}\left( x\right) \int
\int_{\left\vert \varepsilon \right\vert \leq \kappa _{n}}\frac{1}{\pi
\left( m\left( u\right) +\varepsilon \right) }K_{h}\left( u-\!x\right)
\varepsilon dB_{n_{0}}\left( T\left( u,\varepsilon \right) \right) , \\
\zeta _{4n_{0}}\left( x\right) &=&h^{1/2}s_{n}^{-1/2}\left( x\right) \int
\int_{\left\vert \varepsilon \right\vert \leq \kappa _{n}}\frac{1}{\pi
\left( m\left( u\right) +\varepsilon \right) }K_{h}\left( u-\!x\right)
\varepsilon dW_{n_{0}}\left( T\left( u,\varepsilon \right) \right) ,
\end{eqnarray*}%
where $s_{n}\left( x\right) =\int_{\left\vert \varepsilon \right\vert \leq
\kappa _{n}}\varepsilon ^{2}f_{X,\varepsilon |\delta =1}\left( x,\varepsilon
\right) d\varepsilon ,$ $B_{n_{0}}\left( T\left( u,\varepsilon \right)
\right) $ is the sequence of Brownian bridges in (\ref%
{EQ:strong_approximation}) and $W_{n_{0}}\left( T\left( u,\varepsilon
\right) \right) $ is the sequence of Wiener process satisfying $%
B_{n_{0}}\!\left( u,s\right) \!=\!W_{n_{0}}\!\left( u,s\right) $ $%
-usW_{n_{0}}\left( 1,1\right) $. Moreover, define
\begin{equation*}
\zeta _{5n_{0}}\left( x\right) =h^{1/2}s_{n}^{-1/2}\left( x\right) \int
s_{n}^{1/2}\left( u\right) K_{h}\left( u-\!x\right) dW\left( u\right) ,
\end{equation*}%
and
\begin{equation*}
\zeta _{6n_{0}}\left( x\right) =h^{1/2}\int K_{h}\left( u-\!x\right)
dW\left( u\right) ,
\end{equation*}%
where $W\left( u\right) $ is a two-sided Wiener process on $\left( -\infty
,+\infty \right) $. Conditional on $\Delta _{n}=n_{0}$, according to Theorem 3.1 in Bickel and
Rosenblatt (1973), one has
\begin{equation}
P\left[ \left. a_{h}\left\{ \sup_{x\in \left[ a_{0},b_{0}\right] }\left\vert
\zeta _{6n_{0}}\left( x\right) \right\vert /\lambda ^{1/2}\left( K\right)
-b_{h}\right\} \leq t\right\vert \Delta _{n}=n_{0} \right] \rightarrow \exp
\left\{ -2\exp \left( -t\right) \right\}  \label{EQ:xi6_distribution}
\end{equation}
$\forall t\in \mathbb{R}$, as $n_0 $ (and thus $n$) $\to \infty$. Here $a_{h},b_{h}$, and $\lambda\left( K\right)$ are given in Theorem 2.

\begin{lemma}
\label{LEM:process_aproxi} Under Assumptions (A1)--(A5), conditional on $%
\Delta _{n}=n_{0}$, for an increasing sequence $\{n_0\}$, one has
\begin{gather*}
(a)\sup_{x\in \left[ a_{0},b_{0}\right] }\left\vert \zeta _{2n_{0}}\left(
x\right) -\zeta _{3n_{0}}\left( x\right) \right\vert =o_{p}\left( \log
^{-1/2}n\right) , \\
(b)\sup_{x\in \left[ a_{0},b_{0}\right] }\left\vert \zeta _{3n_{0}}\left(
x\right) -\zeta _{4n_{0}}\left( x\right) \right\vert =o_{p}\left( \log
^{-1/2}n\right) , \\
(c)\sup_{x\in \left[ a_{0},b_{0}\right] }\left\vert \zeta _{5n_{0}}\left(
x\right) -\zeta _{6n_{0}}\left( x\right) \right\vert =o_{p}\left( \log
^{-1/2}n\right) ,
\end{gather*}
\end{lemma}
\noindent
as $n_0 \to \infty $.

\noindent \textbf{Proof of Lemma \ref{LEM:process_aproxi}(a). }By
integration by parts, one has {\
\begin{align*}
& \zeta _{2n_{0}}\left( x\right) -\zeta _{3n_{0}}\left( x\right) \\
& =h^{1/2}\left\{ s_{n}\left( x\right) \right\} ^{-1/2}\!\!\!\int \int_{\left\vert
\varepsilon \right\vert \leq \kappa _{n}}\!\!\!\frac{K_{h}\left( u-\!x\right)
\varepsilon }{\pi \left( m\left( u\right) +\varepsilon \right) }d\left\{
Z_{n_{0}}\left( u,\varepsilon \right)\!\!-\!\!B_{n_{0}}\!\!\left( T\left( u,\varepsilon
\right) \right) \right\} \\
& =h^{-1/2}\left\{ s_{n}\left( x\right) \right\}
^{-1/2}\int_{-1}^{1}\int_{\left\vert \varepsilon \right\vert \leq \kappa
_{n}}\frac{K\left( v\right) \varepsilon }{\pi \left( m\left( x+hv\right)
+\varepsilon \right) }d\left\{ Z_{n_{0}}\left( x+hv,\varepsilon \right)
\!-\!B_{n_{0}}\!\left( T\left( x+hv,\varepsilon \right) \right) \right\} \\
& =h^{-1/2}\left\{ s_{n}\left( x\right) \right\}
^{-1/2}\!\!\int_{-1}^{1}\int_{\left\vert \varepsilon \right\vert \leq \kappa
_{n}}\!\!\left\{ Z_{n_{0}}\left( x+hv,\varepsilon \right)\!-\!B_{n_{0}}\!\left(
T\left( x+hv,\varepsilon \right) \right) \right\} d\frac{K\left( v\right)
\varepsilon }{\pi \left( m\left( x+hv\right) +\varepsilon \right) } \\
& \hspace{0.4cm}+h^{-1/2}\left\{ s_{n}\left( x\right) \right\}
^{-1/2}\int_{-1}^{1}\left\{ Z_{n_{0}}\left( x+hv,\kappa _{n}\right)
\!-\!B_{n_{0}}\left( T\left( x+hv,\kappa _{n}\right) \right) \right\} d\frac{%
K\left( v\right) \kappa _{n}}{\pi \left( m\left( x+hv\right) +\kappa
_{n}\right) } \\
& \hspace{0.4cm}+h^{-1/2}\left\{ s_{n}\left( x\right) \right\}
^{-1/2}\!\int_{-1}^{1}\left\{ Z_{n_{0}}\left( x+hv,-\kappa _{n}\right)
\!-\!B_{n_{0}}\left( T\left( x+hv,-\kappa _{n}\right) \right) \right\} d\frac{%
K\left( v\right) \kappa _{n}}{\pi \left( m\left( x\!+\!hv\right) \!-\!\kappa
_{n}\right) }.
\end{align*}%
For the first term, by (\ref{EQ:strong_approximation}) and (\ref{EQ:kn}),
one has
\begin{gather*}
\sup_{x\in \left[ a_{0},b_{0}\right] }\left\vert h^{-1/2}\left\{ s_{n}\left(
x\right) \right\} ^{-1/2}\int_{-1}^{1}\int_{\left\vert \varepsilon
\right\vert \leq \kappa _{n}}\left\{ Z_{n_{0}}\left( x+hv,\varepsilon
\right) -B_{n_{0}}\left( T\left( x+hv,\varepsilon \right) \right) \right\}
\right. \\
\left. d\frac{K\left( v\right) \varepsilon }{\pi \left( m\left( x+hv\right)
+\varepsilon \right) }\right\vert =O_{p}\left( \kappa
_{n}^{2}n^{-1/2}h^{-1/2}\log ^{2}n\right) =o_{p}\left( \log ^{-1/2}n\right) .
\end{gather*}%
Similarly,
\begin{gather*}
\sup_{x\in \left[ a_{0},b_{0}\right] }\left\vert h^{-1/2}\left\{ s_{n}\left(
x\right) \right\} ^{-1/2}\int_{-1}^{1}\left\{ Z_{n_{0}}\left( x+hv,\kappa
_{n}\right) -B_{n_{0}}\left( T\left( x+hv,\kappa _{n}\right) \right)
\right\} \right. \\
\left. d\frac{K\left( v\right) \kappa _{n}}{\pi \left( m\left( x+hv\right)
+\kappa _{n}\right) }\right\vert =O_{p}\left( \kappa
_{n}n^{-1/2}h^{-1/2}\log ^{2}n\right) =o_{p}\left( \log ^{-1/2}n\right),
\end{gather*}%
and
\begin{gather*}
\sup_{x\in \left[ a_{0},b_{0}\right] }\left\vert h^{-1/2}\left\{ s_{n}\left(
x\right) \right\} ^{-1/2}\int_{-1}^{1}\left\{ Z_{n_{0}}\left( x+hv,-\kappa
_{n}\right) -B_{n_{0}}\left( T\left( x+hv,-\kappa _{n}\right) \right)
\right\} \right. \\
\left. d\frac{K\left( v\right) \kappa _{n}}{\pi \left( m\left( x+hv\right)
-\kappa _{n}\right) }\right\vert =O_{p}\left( \kappa
_{n}n^{-1/2}h^{-1/2}\log ^{2}n\right) =o_{p}\left( \log ^{-1/2}n\right) ,
\end{gather*}%
completing the proof. }

\noindent {\textbf{Proof of Lemma \ref{LEM:process_aproxi}(b).}} It is
clear that the Jacobian of the transformation $T$ is $f_{X,\varepsilon
|\delta =1}(x,\varepsilon )$. Then one has
\begin{align*}
& \left\vert \zeta _{3n_{0}}\left( x\right) -\zeta _{4n_{0}}\left( x\right)
\right\vert \\
& =\left\vert h^{1/2}\left\{ s_{n}\left( x\right) \right\} ^{-1/2}\!\!\int\!
\int_{\left\vert \varepsilon \right\vert \leq \kappa _{n}}\!\!\frac{1}{\pi
\left( m\left( u\right) +\varepsilon \right) }K_{h}\left( u-\!x\right)
\varepsilon d\left\{ B_{n_{0}}\left( T\left( u,\varepsilon \right) \right)
\!-\!W_{n_{0}}\left( T\left( u,\varepsilon \right) \right) \right\}
\right\vert \\
& =\left\vert h^{-1/2}\left\{ s_{n}\left( x\right) \right\} ^{-1/2}\!\!\int
\!\int_{\left\vert \varepsilon \right\vert \leq \kappa _{n}}\!\!\frac{1}{\pi
\left( m\left( u\right) +\varepsilon \right) }K\left( \frac{u-\!x}{h}\right)
\varepsilon f_{X,\varepsilon |\delta =1}\left( u,\varepsilon \right)
dud\varepsilon \right\vert \left\vert W_{n_{0}}\left( 1,1\right) \right\vert
\\
& =\left\vert h^{1/2}\left\{ s_{n}\left( x\right) \right\}
^{-1/2}\!\int_{\left\vert \varepsilon \right\vert \leq \kappa _{n}}\frac{1}{%
\pi \left( m\left( x\right) +\varepsilon \right) }\varepsilon
f_{X,\varepsilon |\delta =1}\left( x,\varepsilon \right) d\varepsilon
\left\{ 1+u\left( 1\right) \right\} \right\vert \left\vert W_{n_{0}}\left(
1,1\right) \right\vert \\
& =U_{p}\left( h^{1/2}\right) =u_{p}\left( \log ^{-1/2}n\right) .
\end{align*}%
The proof is completed.

\noindent {\textbf{Proof of Lemma \ref{LEM:process_aproxi}(c). }} Note
that conditional on $\Delta _{n}=n_{0}$,
\begin{align*}
& \left\vert \zeta _{5n_{0}}\left( x\right) -\zeta _{6n_{0}}\left( x\right)
\right\vert \\
& =\left\vert h^{1/2}\int \left[ \left\{ s_{n}\left( u\right) \right\}
^{1/2}\left\{ s_{n}\left( x\right) \right\} ^{-1/2}-1\right] K_{h}\left(
u-\!x\right) dW\left( u\right) \right\vert \\
& =\left\vert h^{-1/2}\int_{-1}^{1}\left[ \left\{ s_{n}\left( x+hv\right)
\right\} ^{1/2}\left\{ s_{n}\left( x\right) \right\} ^{-1/2}-1\right]
K\left( v\right) dW\left( x+hv\right) \right\vert \\
& =\left\vert -h^{-1/2}\int_{-1}^{1}W\left( x+hv\right) \frac{\partial }{%
\partial v}\left\{ \left( s_{n}^{1/2}\left( x+hv\right) s_{n}^{-1/2}\left(
x\right) -1\right) K\left( v\right) \right\} dv\right\vert .
\end{align*}%
%
%
%
%
%
%
%
%
%
%
%
%
It is readily seen that $\sup_{x\in \left[ a_{0},b_{0}\right] ,v\in \left[
-1,1\right] }\left\vert \frac{\partial s_{n}^{1/2}\left( x+hv\right) }{%
\partial v}\right\vert $ $=O_{p}\left( h\right) .$ Therefore,%
\begin{equation*}
\sup_{x\in \left[ a_{0},b_{0}\right] }\left\vert \zeta _{5n_{0}}\left(
x\right) -\zeta _{6n_{0}}\left( x\right) \right\vert =O_{p}\left(
h^{1/2}\right) =o_{p}\left( \log ^{-1/2}n\right) ,
\end{equation*}%
completing the proof.

\begin{lemma}
\label{LEM:xi4_5_same} Conditional on $\Delta _{n}=n_{0}$ for an increasing sequence $\{n_0\}$,  the stochastic
processes $\zeta _{4n_{0}}\left( x\right) $ and $\zeta _{5n_{0}}\left(
x\right) $ have the same asymptotic distribution as $n_0 \to \infty $.
\end{lemma}

\noindent {\textbf{Proof of Lemma \ref{LEM:xi4_5_same}. }}Clearly,
conditional on $\Delta _{n}=n_{0}$, $\zeta _{4n_{0}}\left( x\right) $ is a
Gaussian process satisfying $\limfunc{E}\left\{ \zeta _{4n_{0}}\left(
x\right) \Big \vert \Delta _{n}=n_{0}\right\} =0$ and
\begin{align*}
& \limfunc{E}\left\{ \zeta _{4n_{0}}\left( x\right) \zeta _{4n_{0}}\left(
x^{\prime }\right) |\Delta _{n}=n_{0} \right\} \!=\!\limfunc{E}\left[ h\left\{
s_{n}\left( x\right) \right\} ^{-1/2}\left\{ s_{n}\left( x^{\prime }\right)
\right\} ^{-1/2}\!\int\!\!\int_{\left\vert \varepsilon \right\vert \leq \kappa
_{n}}\!\!\frac{1}{\pi \left( m\left( u\right) +\varepsilon \right) }
\right. \\
& \hspace{0.5cm}\left. \times K_{h}\left( u-\!x\right) \varepsilon dW_{n_{0}}\left(
T\left( u,\varepsilon \right) \right) \int \int_{\left\vert \varepsilon
\right\vert \leq \kappa _{n}}\frac{1}{\pi \left( m\left( u\right)
+\varepsilon \right) }K_{h}\left( u-\!x^{\prime }\right) \varepsilon
dW_{n_{0}}\left( T\left( u,\varepsilon \right) \right) \right] \\
& =h\left\{ s_{n}\left( x\right) \right\} ^{-1/2}\left\{ s_{n}\left(
x^{\prime }\right) \right\} ^{-1/2}\int \int_{\left\vert \varepsilon
\right\vert \leq \kappa _{n}}\frac{\varepsilon ^{2}}{\pi ^{2}\left( m\left(
u\right) +\varepsilon \right) }K_{h}\left( u-\!x\right) K_{h}\left(
u-\!x^{\prime }\right) \times \\
& \hspace{8cm}f_{X,\varepsilon |\delta =1}\left( u,\varepsilon \right)
dud\varepsilon \\
& =h\left\{ s_{n}\left( x\right) \right\} ^{-1/2}\left\{ s_{n}\left(
x^{\prime }\right) \right\} ^{-1/2}\int K_{h}\left( u-\!x\right) K_{h}\left(
u-\!x^{\prime }\right) s_{n}\left( u\right) du.
\end{align*}
Next, notice that conditional on $\Delta _{n}=n_{0}$, $\zeta _{5n_{0}}\left(
x\right) $ is also a Gaussian process with mean $0$ and covariance function $%
\limfunc{E}\left\{ \zeta _{5n_{0}}\left( x\right) \zeta _{5n_{0}}\left(
x^{\prime }\right) \Big \vert \Delta _{n}=n_{0}\right\} $ equal to
\begin{align*}
& h\left\{ s_{n}\left( x\right) \right\} ^{-1/2}\left\{ s_{n}\left(
x^{\prime }\right) \right\} ^{-1/2}\times \\
& \hspace{2cm}\limfunc{E}\left[ \int \left\{ s_{n}\left( u\right) \right\}
^{1/2}K_{h}\left( u-\!x\right) dW\left( u\right) \int \left\{ s_{n}\left(
u\right) \right\} ^{1/2}K_{h}\left( u-\!x^{\prime }\right) dW\left( u\right) %
\right] \\
& =h\left\{ s_{n}\left( x\right) \right\} ^{-1/2}\left\{ s_{n}\left(
x^{\prime }\right) \right\} ^{-1/2}\int K_{h}\left( u-\!x\right) K_{h}\left(
u-\!x^{\prime }\right) s_{n}\left( u\right) du \\
& =\limfunc{E}\left\{ \zeta _{4n_{0}}\left( x\right) \zeta _{4n_{0}}\left(
x^{\prime }\right) \Big \vert \Delta _{n}=n_{0} \right\} .
\end{align*}%
Thus, $\zeta _{4n_{0}}\left( x\right) $ and $\zeta _{5n_{0}}\left( x\right) $
have the same asymptotic distribution which completes the proof.

Lemmas \ref{LEM:process_aproxi} and \ref{LEM:xi4_5_same}, expression (\ref%
{EQ:xi6_distribution}), and Slutsky's Theorem imply that
\begin{equation}
P\left[ \left. a_{h}\left\{ \sup_{x\in \left[ a_{0},b_{0}\right] }\left\vert
\zeta _{2n_{0}}\left( x\right) \right\vert /\lambda ^{1/2}\left( K\right)
-b_{h}\right\} \leq t\right\vert \Delta _{n}=n_{0} \right] \rightarrow \exp
\left\{ -2\exp \left( -t\right) \right\}  \label{EQ:Xi2_distribution}
\end{equation}
$\forall t\in \mathbb{R}$, as $n_0 \to \infty$.

\begin{lemma}
\label{LEM:xi1_xi2_approx_} Under Assumptions (A1)--(A5), conditional on $%
\Delta _{n}=n_{0}$ for an increasing sequence $\{n_0\}$,  one has%
\begin{equation*}
\sup_{x\in \left[ a_{0},b_{0}\right] }\left\vert \zeta _{1n_{0}}\left(
x\right) -\zeta _{2n_{0}}\left( x\right) \right\vert =o_{p}\left( \log
^{-1/2}n\right)
\end{equation*}
as $n_0 \to \infty $.
\end{lemma}

\noindent {\textbf{Proof of Lemma \ref{LEM:xi1_xi2_approx_}. }} Define
\begin{equation*}
\zeta _{1n_{0}}^{\ast }\left( x\right) =h^{1/2}s^{-1/2}\left( x\right) \int
\int_{\left\vert \varepsilon \right\vert \leq \kappa _{n}}\frac{1}{\pi
\left( m\left( u\right) +\varepsilon \right) }K_{h}\left( u-\!x\right)
\varepsilon dZ_{n_{0}}\left( u,\varepsilon \right) .
\end{equation*}%
To prove the lemma, it is sufficient to prove that conditional on $%
\Delta _{n}=n_{0}$
\begin{equation}
\sup_{x\in \left[ a_{0},b_{0}\right] }\left\vert \zeta _{1n_{0}}\left(
x\right) -\zeta _{1n_{0}}^{\ast }\left( x\right) \right\vert =o_{p}\left(
\log ^{-1/2}n\right)  \label{EQ:xi1_xi1star}
\end{equation}%
and
\begin{equation}
\sup_{x\in \left[ a_{0},b_{0}\right] }\left\vert \zeta _{2n_{0}}\left(
x\right) -\zeta _{1n_{0}}^{\ast }\left( x\right) \right\vert =o_{p}\left(
\log ^{-1/2}n\right)  \label{EQ:xi1star_xi2}
\end{equation}%
as $n_0 \to \infty$. In the following, we first show (\ref{EQ:xi1star_xi2}).
By the definitions of $s\left( x\right) $, $s_{n}\left( x\right) $, and (\ref%
{EQ:kn}), one has
\begin{eqnarray*}
\sup_{x\in \left[ a_{0},b_{0}\right] }\left\vert s\left( x\right)
-s_{n}\left( x\right) \right\vert &=&\sup_{x\in \left[ a_{0},b_{0}\right]
}\tint_{\left\vert \varepsilon \right\vert >\kappa _{n}}\frac{\varepsilon
^{2}}{\pi ^{2}\left( m\left( x\right) +\varepsilon \right) }f_{X,\varepsilon
|\delta =1}\left( x,\varepsilon \right) d\varepsilon \\
&&\hspace{-2.5cm}\leq c_{\pi }^{-2}\sup_{x\in \left[ a_{0},b_{0}\right]
}\tint_{\left\vert \varepsilon \right\vert >\kappa _{n}}\varepsilon
^{2}f_{X,\varepsilon |\delta =1}\left( x,\varepsilon \right) d\varepsilon \\
&&\hspace{-2.5cm}=c_{\pi }^{-2}\sup_{x\in \left[ a_{0},b_{0}\right]
}\tint_{\left\vert \varepsilon \right\vert >\kappa _{n}}\varepsilon
^{2}f_{\varepsilon |X,\delta =1}\left( \varepsilon |x\right) f_{X|\delta
=1}\left( x\right) d\varepsilon \\
&&\hspace{-2.5cm}\leq c_{\pi }^{-2}\kappa _{n}^{-\eta }\sup_{x\in \left[
a_{0},b_{0}\right] }\left\vert f_{X|\delta =1}\left( x\right) \right\vert
M_{\eta }/P(\delta =1)=O\left( h^{2}\log ^{-1}n\right) .
\end{eqnarray*}%
By (\ref{EQ:Xi2_distribution}) and the fact that $b_{h}=O\left( \log
^{1/2}n\right) $, one has $\sup_{x\in \left[ a_{0},b_{0}\right] }$ $%
\left\vert \zeta _{2n_{0}}\left( x\right) \right\vert $ $=O_{p}\left( \log
^{1/2}n\right) $ which implies that%
\begin{eqnarray*}
\sup_{x\in \left[ a_{0},b_{0}\right] }\left\vert \zeta _{2n_{0}}\left(
x\right) -\zeta _{1n_{0}}^{\ast }\left( x\right) \right\vert &=&\sup_{x\in %
\left[ a_{0},b_{0}\right] }\left\vert h^{1/2}\left\{ s^{-1/2}\left( x\right)
-s_{n}^{-1/2}\left( x\right) \right\} \times \right. \\
&&\left. \int \int_{\left\vert \varepsilon \right\vert \leq \kappa _{n}}%
\frac{1}{\pi \left( m\left( u\right) +\varepsilon \right) }K_{h}\left(
u-\!x\right) \varepsilon dZ_{n_{0}}\left( u,\varepsilon \right) \right\vert
\\
&=&O_{p}\left( h^{2}\log ^{-1/2}n\right) =o_{p}\left( \log ^{-1/2}n\right) .
\end{eqnarray*}

We next prove (\ref{EQ:xi1_xi1star}). Notice that
\begin{align*}
&\zeta _{1n_{0}}\left( x\right) -\zeta _{1n_{0}}^{\ast }\left( x\right)
=h^{1/2}s^{-1/2}\left( x\right) \int \int_{\left\vert
\varepsilon \right\vert >\kappa _{n}}\frac{1}{\pi \left( m\left( u\right)
+\varepsilon \right) }K_{h}\left( u-\!x\right) \varepsilon dZ_{n_{0}}\left(
u,\varepsilon \right) \\
&\hspace{1.8cm}\!\!=\!\!s^{-1/2}\left( x\right) \sum\limits_{i=1}^{n_{0}}\left(
n_{0}^{-1}h\right) ^{1/2}\left[ \frac{1}{\pi \left( m\left( X_{i}\right)
+\varepsilon _{i}\right) }K_{h}\left( X_{i}-\!x\right) \varepsilon
_{i}I\left\{ \left\vert \varepsilon _{i}\right\vert >\kappa _{n}\right\}
\right. \\
&\hspace{2.2cm}\!\!\!\!\left. -\limfunc{E}\left\{ \left. \frac{1}{\pi \left( m\left(
X_{i}\right) +\varepsilon _{i}\right) }K_{h}\left( X_{i}-\!x\right)
\varepsilon _{i}I\left\{ \left\vert \varepsilon _{i}\right\vert >\kappa
_{n}\right\} \right\vert \delta _{i}=1\right\} \right] .
\end{align*}%
For convenience, we denote
\begin{eqnarray*}
\varsigma _{i,n}\left( x\right) &=&\left( n_{0}^{-1}h\right)
^{1/2}\log ^{1/2}n\left[ \frac{1}{\pi \left( m\left( X_{i}\right)
+\varepsilon _{i}\right) }K_{h}\left( X_{i}-\!x\right) \varepsilon
_{i}I\left\{ \left\vert \varepsilon _{i}\right\vert >\kappa _{n}\right\}
\right. \\
&\!\!\!\!&\left. -\limfunc{E}\left\{ \frac{1}{\pi \left( m\left(
X_{i}\right) +\varepsilon _{i}\right) }K_{h}\left( X_{i}-\!x\right)
\varepsilon _{i}I\left\{ \left\vert \varepsilon _{i}\right\vert >\kappa
_{n}\right\} \bigg \vert  \delta _{i}=1\right\} \right] .
\end{eqnarray*}%
To prove (\ref{EQ:xi1_xi1star}), it is sufficient to verify that
\begin{equation*}
\sup_{x\in \left[ a_{0},b_{0}\right] }\left\vert
\tsum\nolimits_{i=1}^{n_{0}}\varsigma _{i,n}\left( x\right) \right\vert
=o_{p}\left( 1\right) .
\end{equation*}%
By Theorem 15.6 in Billingsley (1968), it suffices to show: (i) conditional
on $\Delta _{n}=n_{0}$, $\tsum\nolimits_{i=1}^{n_{0}}\varsigma _{i,n}\left(
x\right) $ $\rightarrow 0$ in probability for any given $x\in \left[
a_{0},b_{0}\right] $ and (ii) the tightness of $\tsum\nolimits_{i=1}^{n_{0}}%
\varsigma _{i,n}\left( x\right) $ conditional on $\Delta _{n}=n_{0}$, using
the following moment condition:
\begin{gather*}
\limfunc{E}\left\{ \left\vert \left( \tsum\limits_{i=1}^{n_{0}}\varsigma
_{i,n}\left( x\right) -\tsum\limits_{i=1}^{n_{0}}\varsigma _{i,n}\left(
x_{1}\right) \right) \!\left( \tsum\limits_{i=1}^{n_{0}}\varsigma
_{i,n}\left( x_{2}\right) -\tsum\limits_{i=1}^{n_{0}}\varsigma _{i,n}\left(
x\right) \right) \right\vert \bigg \vert  \Delta _{n}=n_{0}\right\} \\
\leq C\left\vert x_{2}-x_{1}\right\vert ^{2}
\end{gather*}%
almost surely for any $x\in \left[ x_{1},x_{2}\right] $ and some constant $%
C>0$.

Firstly, note that $\varsigma _{i,n}\left( x\right) ,1\leq i\leq n$, are
independent variables with $\limfunc{E}\{\varsigma _{i,n}\left( x\right)
\mid \delta _{i}=1\}=0$ and%
\begin{eqnarray*}
&\!\!\!\!&\limfunc{var}\{\varsigma _{i,n}\left( x\right) \big \vert\delta _{i}=1\}=%
\limfunc{E}\{ \varsigma _{i,n}^{2}\left( x\right) \big \vert\delta _{i}=1\} \\
&\!\!\!\!\leq &n_{0}^{-1}h\left( \log n\right) \limfunc{E}\left[
\frac{1}{\pi ^{2}\left( m\left( X_{i}\right) +\varepsilon _{i}\right) }%
K_{h}^{2}\left( X_{i}-\!x\right) \varepsilon _{i}^{2}I\left\{ \left\vert
\varepsilon _{i}\right\vert >\kappa _{n}\right\} \bigg \vert \delta _{i}=1\right] \\
&\!\!\!\!=&n_{0}^{-1}h\left( \log n\right) \int \int_{\left\vert
\varepsilon \right\vert >\kappa _{n}}\frac{1}{\pi ^{2}\left( m\left(
u\right) +\varepsilon \right) }K_{h}^{2}\left( u-\!x\right) \varepsilon
^{2}f_{X,\varepsilon |\delta =1}\left( u,\varepsilon \right) dud\varepsilon
\\
&\!\!\!\!=&n_{0}^{-1}\left( \log n\right) \int \int_{\left\vert
\varepsilon \right\vert >\kappa _{n}}\frac{1}{\pi ^{2}\left( m\left(
x\right) +\varepsilon \right) }K^{2}\left( v\right) \varepsilon
^{2}f_{X,\varepsilon |\delta =1}\left( x,\varepsilon \right) dvd\varepsilon
\left\{ 1+u\left( 1\right) \right\} \\
&\!\!\!\!\leq &n_{0}^{-1}\left( \log n\right) \int K^{2}\left(
v\right) dv\int_{\left\vert \varepsilon \right\vert >\kappa _{n}}\frac{1}{%
\pi ^{2}\left( m\left( x\right) +\varepsilon \right) }\varepsilon
^{2}f_{X,\varepsilon |\delta =1}\left( x,\varepsilon \right) d\varepsilon
\left\{ 1+u\left( 1\right) \right\} .
\end{eqnarray*}%
Thus, by (\ref{EQ:kn}), one has $\limfunc{var}\{\tsum\nolimits_{i=1}^{n_{0}}%
\varsigma _{i,n}\left( x\right) \mid \Delta _{n}=n_{0}\}=n_{0}\limfunc{var}%
\{\varsigma _{i,n}\left( x\right) $ $\mid \delta _{i}=1\}\rightarrow 0$
which together with Markov's inequality concludes that for any given $x\in %
\left[ a_{0},b_{0}\right] $,
\begin{equation*}
\tsum\nolimits_{i=1}^{n_{0}}\varsigma _{i,n}\left( x\right) \rightarrow 0%
\text{ in probability.}
\end{equation*}

Secondly notice that
\begin{align*}
& \limfunc{E}\left\{\left. \left( \tsum\limits_{i=1}^{n_{0}}\varsigma _{i,n}\left(
x\right) -\tsum\limits_{i=1}^{n_{0}}\varsigma _{i,n}\left( x_{1}\right)
\right) ^{2} \right\vert \Delta _{n}=n_{0}\right\} \\
& =n_{0}^{-1}h\left( \log n\right) \tsum\limits_{i=1}^{n_{0}}\limfunc{E}%
\left[ \left\{ \frac{1}{\pi \left( m\left( X_{i}\right) +\varepsilon
_{i}\right) }\left( K_{h}\left( X_{i}-x\right) \!-\!K_{h}\left(
X_{i}-\!x_{1}\right) \right) \varepsilon _{i}I\left\{ \left\vert \varepsilon
_{i}\right\vert >\kappa _{n}\right\} \right. \right. \\
& \left. \left. -\!\limfunc{E}\!\!\left. \left.  \left( \frac{1}{\pi \left( m\left(
X_{i}\right) +\varepsilon _{i}\right) }\left\{ K_{h}\left( X_{i}-x\right)
\!-\!K_{h}\left( X_{i}-\!x_{1}\right) \right\} \varepsilon _{i}I\left(
\left\vert \varepsilon _{i}\right\vert >\kappa _{n}\right) \left. \right)\right\vert \delta
_{i}=1\right) \right\} ^{2}\right\vert \! \Delta _{n}\!=\!n_{0}\right] .
\end{align*}%
Since $K\left( u\right) $ $\in C^{\left( 1\right) }\left[ -1,1\right] $ by
Assumption (A3),
\begin{align*}
& \limfunc{E}\left\{ \left. \left( \tsum\limits_{i=1}^{n_{0}}\varsigma _{i,n}\left(
x\right) -\tsum\limits_{i=1}^{n_{0}}\varsigma _{i,n}\left( x_{1}\right)
\right) ^{2}\right\vert \Delta _{n}=n_{0}\right\} \\
& \leq C_{1}\left( x-x_{1}\right) ^{2}h^{-2}\left( \log n\right)
\int_{\left\vert \varepsilon \right\vert >\kappa _{n}}\frac{1}{\pi
^{2}\left( m\left( x\right) +\varepsilon \right) }\varepsilon
^{2}f_{X,\varepsilon |\delta =1}\left( x,\varepsilon \right) d\varepsilon
\text{ a.s.}
\end{align*}%
and
\begin{align*}
& \limfunc{E}\left\{ \left. \left( \tsum\limits_{i=1}^{n_{0}}\varsigma _{i,n}\left(
x_{2}\right) -\tsum\limits_{i=1}^{n_{0}}\varsigma _{i,n}\left( x\right)
\right) ^{2}\right\vert \Delta _{n}=n_{0}\right\} \\
& \leq C_{1}\left( x_{2}-x\right) ^{2}h^{-2}\log n\int_{\left\vert
\varepsilon \right\vert >\kappa _{n}}\frac{1}{\pi ^{2}\left( m\left(
x\right) +\varepsilon \right) }\varepsilon ^{2}f_{X,\varepsilon |\delta
=1}\left( x,\varepsilon \right) d\varepsilon \text{ a.s.}
\end{align*}%
for some constant $C_{1}>0$. Therefore, by the Schwarz inequality, one has
that%
\begin{align*}
& \limfunc{E}\left\{ \left\vert \left( \tsum\limits_{i=1}^{n_{0}}\varsigma
_{i,n}\left( x\right) -\tsum\limits_{i=1}^{n_{0}}\varsigma _{i,n}\left(
x_{1}\right) \right) \left. \left( \tsum\limits_{i=1}^{n_{0}}\varsigma _{i,n}\left(
x_{2}\right) -\tsum\limits_{i=1}^{n_{0}}\varsigma _{i,n}\left( x\right)
\right)  \right\vert  \right \vert \Delta _{n}=n_{0}\right\} \\
& \leq \left[ \limfunc{E}\left\{\left. \left( \tsum\limits_{i=1}^{n_{0}}\varsigma
_{i,n}\left( x\right) -\tsum\limits_{i=1}^{n_{0}}\varsigma _{i,n}\left(
x_{1}\right) \right) ^{2} \right\vert \Delta _{n}=n_{0}\right\} \right] ^{1/2} \times \\
& \hspace{30mm} \left[
\limfunc{E}\left\{\left. \left( \tsum\limits_{i=1}^{n_{0}}\varsigma _{i,n}\left(
x_{2}\right) -\tsum\limits_{i=1}^{n_{0}}\varsigma _{i,n}\left( x\right)
\right) ^{2} \right \vert \Delta _{n}=n_{0}\right\} \right] ^{1/2}\\
& \leq C_{1}\left\vert x-x_{1}\right\vert \left\vert x_{2}-x\right\vert
h^{-2}\log n\int_{\left\vert \varepsilon \right\vert >\kappa _{n}}\frac{1}{%
\pi ^{2}\left( m\left( x\right) +\varepsilon \right) }\varepsilon
^{2}f_{X,\varepsilon |\delta =1}\left( x,\varepsilon \right) d\varepsilon
\text{ a.s.}
\end{align*}%
which together with (\ref{EQ:kn}) concludes that
\begin{gather*}
\limfunc{E}\left.\left\{ \left\vert \left( \tsum\limits_{i=1}^{n_{0}}\varsigma
_{i,n}\left( x\right) -\tsum\limits_{i=1}^{n_{0}}\varsigma _{i,n}\left(
x_{1}\right) \right) \!\left( \tsum\limits_{i=1}^{n_{0}}\varsigma
_{i,n}\left( x_{2}\right) -\tsum\limits_{i=1}^{n_{0}}\varsigma _{i,n}\left(
x\right) \right) \right\vert  \right\vert  \Delta _{n}=n_{0}\right\} \\
\leq C^{^{\prime
}}\left\vert x_{2}-x_{1}\right\vert ^{2}\text{ a.s.}
\end{gather*}%
for some $C^{\prime }>0$ verifying the tightness. The proof is completed.


By the definitions of $%
R_{n}\left( x\right) $ in Theorem 1 and $\zeta _{1n_{0}}(x)$
in (\ref{DEF:zeta_1n0}), one has $\zeta
_{1n_{0}}\left( x\right) =\left( nh\right) ^{1/2}r_{n}^{-1/2}s^{-1/2}\left(
x\right) f_{X}\left( x\right) R_{n}\left( x\right) $ given $\Delta _{n}=n_{0}$. This together with
Lemma \ref{LEM:xi1_xi2_approx_}, expression (\ref{EQ:Xi2_distribution}), and
Slutsky's Theorem concludes the extreme value distribution of $R_{n}(x)$ conditional on $\Delta _{n}=n_0$ as stated in the following theorem.
\begin{theorem}
\label{THM:Cond_R_n} Under Assumptions (A1)--(A5), one has that, for any $t\in \mathbb{R}$, as $%
n_{0}\rightarrow \infty ,$
\begin{gather}
P\left[ a_{h}\left\{ \sup_{x\in \left[ a_{0},b_{0}\right] }\left\vert \left(
nh\right) ^{1/2}r_{n}^{-1/2}R_{n}\left( x\right) /d^{1/2}\left( x\right)
\right\vert -b_{h}\right\} \leq t\bigg \vert \Delta _{n}=n_{0}\right]  \notag \\
\rightarrow \exp \left\{ -2\exp \left( -t\right) \right\} .
\label{EQ:Rnmaximaldistribution}
\end{gather}
\end{theorem}

\vskip 2mm \noindent {\large \textbf{ A.2 Proofs of the theorems in Section 2 }}

\noindent {\textbf{Proof of Theorem \ref{THM:mhatpi_m}.} By Lemma S.3 in the Supplementary Materials 
 and Assumption (A5), 
one has }
\begin{equation*}
\mathbf{X}^{T}\mathbf{WX=}\left(
\begin{array}{cc}
L_{n,0}\left( x\right) & L_{n,1}\left( x\right) \\
L_{n,1}\left( x\right) & L_{n,2}\left( x\right)%
\end{array}%
\right) \!=\!f_{X}\left( x\right) \left(
\begin{array}{cc}
1+u_p(h) &U_p(h^{2})\\
U_p(h^{2}) & h^{2} \mu _{2}\left( K\right)+u_p(h^3)
\end{array}
\right)
\end{equation*}%
which implies that
\begin{equation*}
\left( \mathbf{X}^{T}\mathbf{WX}\right) ^{-1}=f_{X}^{-1}\left( x\right)
\left(
\begin{array}{cc}
1+u_{p}\left( h\right) & U_{p}\left(1\right) \\
U_{p}\left( 1\right) & h^{-2}\mu _{2}^{-1}\left( K\right)+u_{p}\left( h^{-1}\right)%
\end{array}%
\right) .
\end{equation*}%
It together with (S.1) 
 and Lemmas S.2 and S.4 
concludes that for any $x\in \left[
a_{0},b_{0}\right]$,
\begin{align*}
& \hat{m}\left( x,\pi \right) -m\left( x\right) \\
& =e_{0}^{T}\left\{ f_{X}^{-1}\left( x\right)
\left(
\begin{array}{cc}
1+u_{p}\left( h\right) & U_{p}\left(1\right) \\
U_{p}\left( 1\right) & h^{-2}\mu _{2}^{-1}\left( K\right)+u_{p}\left( h^{-1}\right)%
\end{array}%
\right)  \right\} \\
& \hspace{3mm}\times \left(
\begin{array}{c}
n^{-1}\tsum\limits_{i=1}^{n}\frac{\delta _{i}}{\pi _{i}}K_{h}\left(
X_{i}-x\right) \varepsilon _{i}+2^{-1}m^{\left( 2\right) }\left( x\right)
f_{X}\left( x\right) \mu _{2}\left( K\right) h^{2}+u_{p}\left( h^{2}\right)
\\
U_{p}\left( n^{-1/2}h^{1/2}\log ^{1/2}n\right)%
\end{array}%
\right) \\
& =R_{n}\left( x\right) +2^{-1}m^{\left( 2\right) }\left( x\right) \mu
_{2}\left( K\right) h^{2}+u_{p}\left( h^{2}\right)+U_p\left(n^{-1/2}h^{1/2}\log
^{1/2}n\right)\\
&= R_{n}\left( x\right) +2^{-1}m^{\left( 2\right) }\left( x\right) \mu
_{2}\left( K\right) h^{2}+u_{p}\left( h^{2}\right).
\end{align*}%
The proof is completed.

\noindent {\textbf{Proof of Theorem \ref{THM:mhatpi_m_distribution}.}
According to Theorem \ref{THM:Cond_R_n}, for any $t\in \mathbb{R}$%
, as $n_{0}\rightarrow \infty $,
\begin{gather*}
P\left[ a_{h}\left. \left\{ \sup_{x\in \left[ a_{0},b_{0}\right] }\left\vert
\left( nh\right) ^{1/2}r_{n}^{-1/2}R_{n}\left( x\right) /d^{1/2}\left(
x\right) \right\vert -b_{h}\right\} \leq t\right\vert \Delta _{n}=n_{0}\right]  \\
\rightarrow \exp \left\{ -2\exp \left( -t\right) \right\} .
\end{gather*}%
Thus one has that for any given $\epsilon >0$ and $t\in \mathbb{R}$, there
exists $N_{0}>0$ such that
\begin{gather*}
\left\vert P\left[ a_{h}\left. \left\{ \sup_{x\in \left[ a_{0},b_{0}\right]
}\left\vert \left( nh\right) ^{1/2}r_{n}^{-1/2}R_{n}\left( x\right)
/d^{1/2}\left( x\right) \right\vert -b_{h}\right\} \leq t\right\vert \Delta
_{n}=n_{0}\right] \right.  \\
\left. -\exp \left\{ -2\exp \left( -t\right) \right\} \right. \Bigg\vert<%
\frac{\epsilon }{2}
\end{gather*}%
for all $n_{0}\geq N_{0}$. On the other hand, since $\Delta
_{n}/n\rightarrow P\left( \delta _{1}=1\right) >0$ a.s., there exists $%
N>N_{0}$ such that when $n\geq N$, $P\left( \Delta _{n}\geq N_{0}\right)
>1-\epsilon /2$. Therefore, unconditional on $\Delta _{n}$, for $n\geq N$,
\begin{gather*}
\left\vert P\left[ a_{h}\left\{ \sup_{x\in \left[ a_{0},b_{0}\right]
}\left\vert \left( nh\right) ^{1/2}r_{n}^{-1/2}R_{n}\left( x\right)
/d^{1/2}\left( x\right) \right\vert -b_{h}\right\} \leq t\right] -\exp
\left\{ -2\exp \left( -t\right) \right\} \right\vert  \\
\leq \sum_{n_{0}=1}^{n}\left\vert P\left[ a_{h}\left. \left\{ \sup_{x\in %
\left[ a_{0},b_{0}\right] }\left\vert \left( nh\right)
^{1/2}r_{n}^{-1/2}R_{n}\left( x\right) /d^{1/2}\left( x\right) \right\vert
-b_{h}\right\} \leq t\right\vert \Delta _{n}=n_{0}\right] \right.  \\
\left. -\exp \left\{ -2\exp \left( -t\right) \right\} \right. \Bigg\vert%
\times P(\Delta _{n}=n_{0})+P(\Delta _{n}=0) \\
\leq \sum_{n_{0}=N_{0}}^{n}\left\vert P\left[ a_{h}\left. \left\{ \sup_{x\in %
\left[ a_{0},b_{0}\right] }\left\vert \left( nh\right)
^{1/2}r_{n}^{-1/2}R_{n}\left( x\right) /d^{1/2}\left( x\right) \right\vert
-b_{h}\right\} \leq t\right\vert \Delta _{n}=n_{0}\right] \right.  \\
\left. -\exp \left\{ -2\exp \left( -t\right) \right\} \right. \Bigg\vert%
\times P(\Delta _{n}=n_{0})+\frac{\epsilon }{2}<\epsilon ,
\end{gather*}%
which together with the fact that the dominating term of $\sup_{x\in \lbrack
a_{0},b_{0}]}\left\vert \hat{m}\left( x,\pi \right) -m\left( x\right)
\right\vert $ is $\sup_{x\in \lbrack a_{0},b_{0}]}\left\vert
R_{n}(x)\right\vert $ in Theorem \ref{THM:mhatpi_m} concludes Theorem \ref%
{THM:mhatpi_m_distribution}. }

\noindent {\textbf{Proof of Theorem }\ref{THM:mhatpi_mhatpihat_oracle}. }By
(S.1) and (S.2) in the Supplementary Materials 
one has
\begin{align*}
& \hat{m}\left( x,\pi \right) -\hat{m}\left( x,\hat{\pi}\right)
=e_{0}^{T}\left(
\begin{array}{cc}
\!L_{n,0}\left( x\right) & L_{n,1}\left( x\right) \! \\
\!L_{n,1}\left( x\right) & L_{n,2}\left( x\right) \!%
\end{array}%
\right) ^{-1}\left(
\begin{array}{c}
\!\!M_{n,0}\left( x\right) \!\!\! \\
\!\!M_{n,1}\left( x\right) \!\!\!%
\end{array}%
\right) \\
& \hspace{1.2cm}-e_{0}^{T}\left(
\begin{array}{cc}
\!\hat{L}_{n,0}\left( x\right) & \hat{L}_{n,1}\left( x\right) \! \\
\!\hat{L}_{n,1}\left( x\right) & \hat{L}_{n,2}\left( x\right) \!%
\end{array}%
\right) ^{-1}\left(
\begin{array}{c}
\!\!\hat{M}_{n,0}\left( x\right) \!\!\! \\
\!\!\hat{M}_{n,1}\left( x\right) \!\!\!%
\end{array}%
\right) .
\end{align*}%
By Lemma S.5, 
it is easily seen that
\begin{equation*}
\sup_{x\in \left[ a_{0},b_{0}\right] }\left\vert \hat{m}\left( x,\pi \right)
-\hat{m}\left( x,\hat{\pi}\right) \right\vert =O_{p}\left( n^{-1/2}\right) ,
\end{equation*}%
completing the proof.

\noindent {\textbf{Proof of Theorem }\ref{THM:duhat_du}. By definition, }%
\begin{eqnarray}
\hat{d}_{n}\left( x\right) &=&\frac{n}{\Delta _{n}}\hat{f}_{X}^{-2}\left(
x\right) \frac h n \dsum\limits_{i=1}^{n}\frac{\delta _{i}}{\hat{\pi}_{i}^{2}%
}K_{h}^{2}\left( X_{i}\!-\!x\right) \hat{\varepsilon}_{i}^{2}.
\label{dhat_form}
\end{eqnarray}%
{Firstly, we study the uniform convergence property of }$\frac h
n\sum\limits_{i=1}^{n}\frac{\delta _{i}}{\hat{\pi}_{i}^{2}}K_{h}^{2}\left(
X_{i}\!-\!x\right) \hat{\varepsilon}_{i}^{2}$. Notice that
\begin{align*}
& \sup_{x\in \left[ a_{0},b_{0}\right] }\left\vert \frac h
n\dsum\limits_{i=1}^{n}\frac{\delta _{i}}{\hat{\pi}_{i}^{2}}K_{h}^{2}\left(
X_{i}\!-\!x\right) \hat{\varepsilon}_{i}^{2}-\frac h n\dsum\limits_{i=1}^{n}%
\frac{\delta _{i}}{\pi _{i}^{2}}K_{h}^{2}\left( X_{i}\!-\!x\right) \hat{%
\varepsilon}_{i}^{2}\right\vert \\
& =\sup_{x\in \left[ a_{0},b_{0}\right] }\left\vert \frac h
n\dsum\limits_{i=1}^{n}\frac{\delta _{i}\left( \pi _{i}^{2}-\hat{\pi}%
_{i}^{2}\right) }{\hat{\pi}_{i}^{2}\pi _{i}^{2}}K_{h}^{2}\left(
X_{i}\!-\!x\right) \left\{ m\left( X_{i}\right) -\hat{m}\left( X_{i},\hat{\pi%
}_{i}\right) +\varepsilon _{i}\right\} ^{2}\right\vert \\
& =o_{p}\left( n^{-1/2}h^{-1}\right) ,
\end{align*}%
and%
\begin{align*}
& \sup_{x\in \left[ a_{0},b_{0}\right] }\left\vert \frac h
n\dsum\limits_{i=1}^{n}\frac{\delta _{i}}{\pi _{i}^{2}}K_{h}^{2}\left(
X_{i}\!-\!x\right) \hat{\varepsilon}_{i}^{2}-\frac h n\dsum\limits_{i=1}^{n}%
\frac{\delta _{i}}{\pi _{i}^{2}}K_{h}^{2}\left( X_{i}\!-\!x\right)
\varepsilon _{i}^{2}\right\vert \\
& =\sup_{x\in \left[ a_{0},b_{0}\right] }\left\vert \frac h
n\dsum\limits_{i=1}^{n}\frac{\delta _{i}}{\pi _{i}^{2}}K_{h}^{2}\left(
X_{i}\!-\!x\right) \left( \hat{\varepsilon}_{i}^{2}-\varepsilon
_{i}^{2}\right) \right\vert \\
& \leq \sup_{x\in \left[ a_{0},b_{0}\right] }\left\vert \frac h
n\dsum\limits_{i=1}^{n}\frac{\delta _{i}}{\pi _{i}^{2}}K_{h}^{2}\left(
X_{i}\!-\!x\right) \left\{ m\left( X_{i}\right) -\hat{m}\left( X_{i},\hat{\pi%
}_{i}\right) \right\} ^{2}\right\vert \\
& \hspace{1cm}+\sup_{x\in \left[ a_{0},b_{0}\right] }\left\vert 2\frac h
n\tsum\limits_{i=1}^{n}\!\frac{\delta _{i}}{\pi _{i}^{2}}K_{h}^{2}\left(
X_{i}\!-\!x\right) \left\{ m\left( X_{i}\right) -\hat{m}\left( X_{i},\hat{\pi%
}_{i}\right) \right\} \varepsilon _{i}\right\vert \\
& =O_{p}\left( n^{-1/2}h^{-3/2}\log ^{1/2}n\right)  ,
\end{align*}%
which imply that
\begin{align}
&\sup_{x\in \left[ a_{0},b_{0}\right] }\left\vert \frac h
n\dsum\limits_{i=1}^{n}\frac{\delta _{i}}{\hat{\pi}_{i}^{2}}K_{h}^{2}\left(
X_{i}\!-\!x\right) \hat{\varepsilon}_{i}^{2}-\frac h n\dsum\limits_{i=1}^{n}%
\frac{\delta _{i}}{\pi _{i}^{2}}K_{h}^{2}\left( X_{i}\!-\!x\right)
\varepsilon _{i}^{2}\right\vert \notag\\
&\hspace{1cm}=O_{p}\left( n^{-1/2}h^{-3/2}\log ^{1/2}n\right).
\label{EQ:pihatkepshat_pikeps}
\end{align}
Secondly, denote $\varepsilon _{i}^{\ast }=\frac{\delta _{i}\varepsilon
_{i}^{2}}{\pi _{i}^{2}}-\limfunc{E}\left( \frac{\delta _{i}\varepsilon
_{i}^{2}}{\pi _{i}^{2}}\Big \vert X_{i}\right) $. By applying the inequality in
Lemma S.1, 
the Borel-Cantelli Lemma, the
truncation and discretization method as in the proof of Lemma S.2, 
one obtains that
\begin{equation}
\sup_{x\in \left[ a_{0},b_{0}\right] }\left\vert \frac h
n\dsum\limits_{i=1}^{n}\!K_{h}^{2}\left( X_{i}\!-x\right) \varepsilon
_{i}^{\ast }\right\vert =O_{p}\left( n^{-1/2}h^{-1/2}\log^{1/2}n\right)  \label{EQ:Kepsstar}
\end{equation}%
as $n\rightarrow \infty $. Meanwhile, similar to the proof of Lemma S.3, 
one can easily show that
\begin{gather}
\sup_{x\in \left[ a_{0},b_{0}\right] }\left\vert \frac h
n\dsum\limits_{i=1}^{n}\!\limfunc{E}\left\{ K_{h}^{2}\left(
X_{i}\!-\!x\right) \frac{\delta _{i}\varepsilon _{i}^{2}}{\pi _{i}^{2}}\mid
X_{i}\right\} -h\limfunc{E}\left\{ K_{h}^{2}\left( X_{1}\!-x\right) \frac{%
\delta _{1}\varepsilon _{1}^{2}}{\pi _{1}^{2}}\right\} \right\vert  \notag \\
=O_{p}\left(  n^{-1/2}h^{-1/2}\log^{1/2}n \right) .  \label{EQ:1}
\end{gather}%
Combining (\ref{EQ:pihatkepshat_pikeps}), (\ref{EQ:Kepsstar}), and (\ref%
{EQ:1}), one has
\begin{equation*}
\sup_{x\in \left[ a_{0},b_{0}\right] }\left\vert \frac h
n\dsum\limits_{i=1}^{n}\!\frac{\delta _{i}}{\hat{\pi}_{i}^{2}}%
K_{h}^{2}\left( X_{i}\!\!-\!\!x\right) \hat{\varepsilon}_{i}^{2}-h\limfunc{E}%
\left\{ \frac{\delta _{1}}{\pi _{1}^{2}}K_{h}^{2}\left( X_{1}\!-x\right)
\varepsilon _{1}^{2}\right\} \right\vert \!=\!O_{p}\left( n^{-1/2}h^{-3/2}\log ^{1/2}n\right) .
\end{equation*}%
Meanwhile, by Lemmas S.3 and S.5, 
one can
easily obtain that
\begin{equation*}
\sup_{x\in \left[ a_{0},b_{0}\right] }\left\vert \hat{f}_{X}\left( x\right)
-f_{X}\left( x\right) \right\vert =o_{p}\left( h_f\right)+O_{p}\left(  n^{-1/2}h_f^{-1/2}\log^{1/2}n \right)=o_{p}\left(  n^{-1/5} \right),
\end{equation*}
since $h_f=O(n^{-1/5})$.
Thus
\begin{align*}
&\sup_{x\in \left[ a_{0},b_{0}\right] }\left\vert \hat{f}_{X}^{-2}\left(
x\right) \frac h n\dsum\limits_{i=1}^{n}\!\frac{\delta _{i}}{\hat{\pi}%
_{i}^{2}}K_{h}^{2}\left( X_{i}\!-\!x\right) \hat{\varepsilon}%
_{i}^{2}-f_{X}^{-2}\left( x\right) h\limfunc{E}\left\{ \frac{\delta _{1}}{%
\pi _{1}^{2}}K_{h}^{2}\left( X_{1}\!-x\right) \varepsilon _{1}^{2}\right\}
\right\vert\\
& \hspace{1cm}=o_{p}\left( n^{-1/5}\right)+ O_{p}\left( n^{-1/2}h^{-3/2}\log ^{1/2}n\right)= O_{p}\left( n^{-1/2}h^{-3/2}\log ^{1/2}n\right) ,
\end{align*}%
which together with the fact that
\begin{align*}
&f_{X}^{-2}\left( x\right) h\limfunc{E}\left\{ \frac{\delta _{1}}{\pi _{1}^{2}%
}K_{h}^{2}\left( X_{1}\!-x\right) \varepsilon _{1}^{2}\right\} \\
&\hspace{1cm}= f_{X}^{-2}\left( x\right) h\limfunc{E}\left\{ \frac{1}{%
\pi _{1}^{2}}K_{h}^{2}\left( X_{1}\!-x\right) \varepsilon _{1}^{2}\Big \vert \delta
_{1}=1\right\} P\left( \delta _{1}=1\right) \\
&\hspace{1cm}=d\left( x\right) P\left( \delta _{1}=1\right)
+u_{p}\left( h\right)
\end{align*}%
implies that
\begin{align}
&\sup_{x\in \left[ a_{0},b_{0}\right] }\left\vert \hat{f}_{X}^{-2}\left(
x\right) \frac h n\dsum\limits_{i=1}^{n}\!\frac{\delta _{i}}{\hat{\pi}%
_{i}^{2}}K_{h}^{2}\left( X_{i}\!-\!x\right) \hat{\varepsilon}%
_{i}^{2}-d\left( x\right) P\left( \delta _{1}=1\right) \right\vert \notag \\
&\hspace{1cm}=O_{p}\left( n^{-1/2}h^{-3/2}\log ^{1/2}n\right).  \label{EQ:dx_pdelta_1}
\end{align}%
It is easily seen from (\ref{dhat_form}), (\ref{limit_for_rn}), and (\ref%
{EQ:dx_pdelta_1}) that%
\begin{align*}
& \sup_{x\in \left[ a_{0},b_{0}\right] }\left\vert \hat{d}_{n}\left(
x\right) -d\left( x\right) \right\vert =O_{p}\left( n^{-1/2}h^{-3/2}\log ^{1/2}n\right) ,
\end{align*}%
completing the proof.

\newpage

\markboth{\hfill{\footnotesize\rm LI CAI, LIJIE GU, QIHUA WANG, AND SUOJIN WANG} \hfill}
{\hfill {\footnotesize\rm SIMULTANEOUS CONFIDENCE BANDS FOR MISSING COVARIATES DATA} \hfill}

\bibhang=1.7pc \bibsep=2pt \fontsize{9}{14pt plus.8pt minus .6pt}%
\selectfont
\renewcommand{\bibname}

\markboth{\hfill{\footnotesize\rm LI CAI, LIJIE GU, QIHUA WANG, AND SUOJIN WANG} \hfill}
{\hfill {\footnotesize\rm SIMULTANEOUS CONFIDENCE BANDS FOR MISSING COVARIATES DATA} \hfill}

\renewcommand{\baselinestretch}{2} 
\vskip .65cm \noindent School of Statistics and
Mathematics, Zhejiang Gongshang University, Hangzhou, 310018, China. \vskip %
0pt \noindent E-mail: caili16@126.com \vskip 2pt \noindent School of
Mathematical Sciences, Soochow University, Suzhou, 215006, China.\vskip 0pt
\noindent E-mail: gulijie@suda.edu.cn \vskip 2pt \noindent School of
Statistics and Mathematics, Zhejiang Gongshang University, Hangzhou, 310018,
China. \vskip 0pt \noindent E-mail: qhwang@amss.ac.cn \vskip 2pt \noindent
Department of Statistics, Texas A\&M University, Texas, TX 77843, U.S.A. %
\vskip 0pt \noindent E-mail: sjwang@stat.tamu.edu

\markboth{\hfill{\footnotesize\rm LI CAI, LIJIE GU, QIHUA WANG, AND SUOJIN WANG} \hfill}
{\hfill {\footnotesize\rm SIMULTANEOUS CONFIDENCE BANDS FOR MISSING COVARIATES DATA} \hfill}

\newpage

\begin{table}[tbp]
\caption{ Empirical coverage frequencies of the SCB in (\protect\ref%
{EQ:feasible_SCB}) and the SCB in the complete case (SCB-CC) from 1000
replications and their corresponding average lengths (inside parentheses)
under the selection probability model (i) with parameters $(\protect\alpha_0,%
\protect\alpha_1)=(1.8,1)$. }
\label{TAB:coverage_logit_2_1}
\begin{center}
\begin{tabular}{ccccccccccccc}
\hline\hline
& \multirow{2}{*}{$n$} & \multirow{2}{*}{$1-\alpha$} & \multicolumn{2}{c}{$%
\mbox{Case 1}$} &  & \multicolumn{2}{c}{$\mbox{Case 2}$} &  &  \\
\cline{4-5}\cline{7-8}
& \multirow{2}{*} &  & $\mbox{SCB}$ & $\mbox{SCB-CC}$ &  & $\mbox{SCB}$ & $%
\mbox{SCB-CC}$ &  &  \\ \hline
& \multirow{2}{*}{400} & $0.95$ & $0.938(1.102)$ & $0.422(0.910)$ &  & $%
0.953(1.070)$ & $0.573(0.910)$ &  &  \\
&  & $0.99$ & $0.993(1.422)$ & $0.832(1.175)$ &  & $0.994(1.380)$ & $%
0.901(1.174)$ &  & \\ \hline
& \multirow{2}{*}{600} & $0.95$ & $0.954(0.934)$ & $0.284(0.774)$ &  & $%
0.955(0.908)$ & $0.443(0.771)$ &  & \\
&  & $0.99$ & $0.999(1.197)$ & $0.705(0.992)$ &  & $0.994(1.164)$ & $%
0.850(0.988)$ &  & \\ \hline
& \multirow{2}{*}{800} & $0.95$ & $0.949(0.833)$ & $0.180(0.690)$ &  & $%
0.949(0.811)$ & $0.349(0.689)$ &  & \\
&  & $0.99$ & $0.998(1.063)$ & $0.596(0.881)$ &  & $0.996(1.035)$ & $%
0.779(0.879)$ &  &  \\ \hline
& \multirow{2}{*}{$n$} & \multirow{2}{*}{$1-\alpha$} & \multicolumn{2}{c}{$%
\mbox{Case 3}$} &  & \multicolumn{2}{c}{$\mbox{Case 4}$} &  & \\
\cline{4-5}\cline{7-8}
&  & \multirow{2}{*} & $\mbox{SCB}$ & $\mbox{SCB-CC}$ &  & $\mbox{SCB}$ & $%
\mbox{SCB-CC}$ &  & \\ \hline
& \multirow{2}{*}{400} & $0.95$ & $0.938(0.945)$ & $0.816(0.870)$ &  & $%
0.942(0.985)$ & $0.789(0.882)$ &  &  \\
&  & $0.99$ & $0.991(1.216)$ & $0.976(1.120)$ &  & $0.995(1.265)$ & $%
0.970(1.133)$ &  & \\ \hline
& \multirow{2}{*}{600} & $0.95$ & $0.932(0.806) $ & $0.800(0.744)$ &  & $%
0.940(0.844)$ & 0.757(0.756) &  & \\
&  & $0.99$ & $0.994(1.031) $ & $0.982(0.951)$ &  & $0.997(1.077)$ &
0.963(0.965) &  & \\ \hline
& \multirow{2}{*}{800} & $0.95$ & $0.934(0.720) $ & $0.768(0.666)$ &  & $%
0.937(0.760)$ & $0.699(0.680)$ &  & \\
&  & $0.99$ & $0.995(0.916) $ & $0.962(0.847)$ &  & $0.997(0.964)$ & $%
0.940(0.863)$ &  & \\ \hline\hline
\end{tabular}%
\end{center}
\end{table}

\begin{table}[tbp]
\caption{ Empirical coverage frequencies of the SCB in (\protect\ref%
{EQ:feasible_SCB}) and the SCB in the complete case (SCB-CC) from 1000
replications and their corresponding average lengths (inside parentheses)
under the selection probability model (i) with parameters $(\protect\alpha_0,%
\protect\alpha_1)=(0.2,0.6)$.}
\label{TAB:coverage_logit_0206}
\begin{center}
\begin{tabular}{ccccccccccccc}
\hline\hline
& \multirow{2}{*}{$n$} & \multirow{2}{*}{$1-\alpha$} & \multicolumn{2}{c}{$%
\mbox{Case 1}$} &  & \multicolumn{2}{c}{$\mbox{Case 2}$} &  & \\
\cline{4-5}\cline{7-8}
& \multirow{2}{*} &  & $\mbox{SCB}$ & $\mbox{SCB-CC}$ &  & $\mbox{SCB}$ & $%
\mbox{SCB-CC}$ &  & \\ \hline
& \multirow{2}{*}{400} & $0.95$ & $0.916(1.303)$ & $0.365(1.118)$ &  & $%
0.922(1.267)$ & $0.402(1.103)$ &  &\\
&  & $0.99$ & $0.994(1.687)$ & $0.774(1.447)$ &  & $0.992(1.641)$ & $%
0.816(1.428)$ &  &   \\ \hline
& \multirow{2}{*}{600} & $0.95$ & $0.938(1.105)$ & $0.219(0.945)$ &  & $%
0.953(1.076)$ & $0.247(0.932)$ &  &   \\
&  & $0.99$ & $0.993(1.422)$ & $0.696(1.215)$ &  & $0.997(1.385)$ & $%
0.741(1.199)$ &  & \\ \hline
& \multirow{3}{*}{800} & $0.95$ & $0.942(0.979)$ & $0.125(0.837)$ &  & $%
0.950(0.957)$ & $0.138(0.829)$ &  &  \\
&  & $0.99$ & $0.993(1.256)$ & $0.562(1.074)$ &  & $0.998(1.227)$ & $%
0.607(1.062)$ &  &  \\ \hline
& \multirow{2}{*}{$n$} & \multirow{2}{*}{$1-\alpha$} & \multicolumn{2}{c}{$%
\mbox{Case 3}$} &  & \multicolumn{2}{c}{$\mbox{Case 4}$} &  &   \\
\cline{4-5}\cline{7-8}
&  & \multirow{2}{*} & $\mbox{SCB}$ & $\mbox{SCB-CC}$ &  & $\mbox{SCB}$ & $%
\mbox{SCB-CC}$ &  & \\ \hline
& \multirow{2}{*}{400} & $0.95$ & $0.918(1.103)$ & $0.525(0.988)$ &  & $%
0.935(1.160)$ & $0.495(1.016)$ &  &  \\
&  & $0.99$ & $0.991(1.425)$ & $0.873(1.277)$ &  & $0.992(1.492)$ & $%
0.857(1.308)$ &  & \\ \hline
& \multirow{2}{*}{600} & $0.95$ & $0.935(0.948) $ & $0.417(0.846)$ &  & $%
0.948(0.999)$ & 0.351(0.870) &  &\\
&  & $0.99$ & $0.996(1.215) $ & $0.836(1.085)$ &  & $0.997(1.277)$ &
0.826(1.112) &  &  \\ \hline
& \multirow{3}{*}{800} & $0.95$ & $0.939(0.845) $ & $0.318(0.758)$ &  & $%
0.936(0.889)$ & $0.248(0.777)$ &  &  \\
&  & $0.99$ & $0.992(1.078) $ & $0.766(0.967)$ &  & $1.000(1.131)$ & $%
0.715(0.989)$ &  & \\ \hline\hline
\end{tabular}%
\end{center}
\end{table}

\begin{table}[tbp]
\caption{ Empirical coverage frequencies of the SCB in (\protect\ref%
{EQ:feasible_SCB}) and the SCB in the complete case (SCB-CC) from 1000
replications and their corresponding average lengths (inside parentheses)
under the selection probability model (ii) with parameters $(\protect\alpha%
_0^{*},\protect\alpha_1^{*})=(1,0.5)$. }
\label{TAB:coverage_probit_06_03}
\begin{center}
\begin{tabular}{ccccccccccccc}
\hline\hline
& \multirow{2}{*}{$n$} & \multirow{2}{*}{$1-\alpha$} & \multicolumn{2}{c}{$%
\mbox{Case 1}$} &  & \multicolumn{2}{c}{$\mbox{Case 2}$} &  & \\
\cline{4-5}\cline{7-8}
& \multirow{2}{*} &  & $\mbox{SCB}$ & $\mbox{SCB-CC}$ &  & $\mbox{SCB}$ & $%
\mbox{SCB-CC}$ &  &  \\ \hline
& \multirow{2}{*}{400} & $0.95$ & $0.944(1.062)$ & $0.527(0.926)$ &  & $%
0.946(1.043)$ & $0.643(0.923)$ &  & \\
&  & $0.99$ & $0.994(1.370)$ & $0.888(1.194)$ &  & $0.996(1.344)$ & $%
0.940(1.191)$ &  &   \\ \hline
& \multirow{2}{*}{600} & $0.95$ & $0.950(0.902)$ & $0.409(0.787)$ &  & $%
0.946(0.887)$ & $0.540(0.784)$ &  &  \\
&  & $0.99$ & $0.996(1.156)$ & $0.819(1.009)$ &  & $1.000(1.136)$ & $%
0.900(1.005)$ &  &  \\ \hline
& \multirow{2}{*}{800} & $0.95$ & $0.952(0.808)$ & $0.330(0.705)$ &  & $%
0.948(0.791)$ & $0.458(0.702)$ &  &  \\
&  & $0.99$ & $0.996(1.030)$ & $0.745(0.899)$ &  & $0.998(1.009)$ & $%
0.856(0.894)$ &  & \\ \hline
& \multirow{2}{*}{$n$} & \multirow{2}{*}{$1-\alpha$} & \multicolumn{2}{c}{$%
\mbox{Case 3}$} &  & \multicolumn{2}{c}{$\mbox{Case 4}$} &  &  \\
\cline{4-5}\cline{7-8}
&  & \multirow{2}{*} & $\mbox{SCB}$ & $\mbox{SCB-CC}$ &  & $\mbox{SCB}$ & $%
\mbox{SCB-CC}$ &  &  \\ \hline
& \multirow{2}{*}{400} & $0.95$ & $0.931(0.939)$ & $0.815(0.872)$ &  & $%
0.941(0.974)$ & $0.789(0.887)$ &  &  \\
&  & $0.99$ & $0.992(1.209)$ & $0.976(1.123)$ &  & $0.995(1.252)$ & $%
0.973(1.140)$ &  & \\ \hline
& \multirow{2}{*}{600} & $0.95$ & $0.928(0.803) $ & $0.786(0.748)$ &  & $%
0.943(0.836)$ & 0.760(0.761) &  &    \\
&  & $0.99$ & $0.996(1.027) $ & $0.979(0.957)$ &  & $0.998(1.066)$ &
0.968(0.972) &  &  \\ \hline
& \multirow{2}{*}{800} & $0.95$ & $0.938(0.718) $ & $0.752(0.670)$ &  & $%
0.934(0.746)$ & $0.702(0.681)$ &  &  \\
&  & $0.99$ & $0.997(0.914) $ & $0.961(0.853)$ &  & $0.997(0.948)$ & $%
0.945(0.865)$ &  & \\ \hline\hline
\end{tabular}%
\end{center}
\end{table}

\begin{table}[tbp]
\caption{ Empirical coverage frequencies of the SCB in (\protect\ref%
{EQ:feasible_SCB}) and the SCB in the complete case (SCB-CC) from 1000
replications and their corresponding average lengths (inside parentheses)
under the selection probability model (ii) with parameters $(\protect\alpha%
_0^{*},\protect\alpha_1^{*})=(0.1,0.3)$. }
\label{TAB:coverage_probit_0502}
\begin{center}
\begin{tabular}{ccccccccccccc}
\hline\hline
& \multirow{2}{*}{$n$} & \multirow{2}{*}{$1-\alpha$} & \multicolumn{2}{c}{$%
\mbox{Case 1}$} &  & \multicolumn{2}{c}{$\mbox{Case 2}$} &  &  \\
\cline{4-5}\cline{7-8}
& \multirow{2}{*} &  & $\mbox{SCB}$ & $\mbox{SCB-CC}$ &  & $\mbox{SCB}$ & $%
\mbox{SCB-CC}$ &  & \\ \hline
& \multirow{2}{*}{400} & $0.95$ & $0.931(1.265)$ & $0.514(1.137)$ &  & $%
0.923(1.226)$ & $0.548(1.114)$ &  &  \\
&  & $0.99$ & $0.995(1.636)$ & $0.872(1.471)$ &  & $0.998(1.588)$ & $%
0.896(1.444)$ &  & \\ \hline
& \multirow{2}{*}{600} & $0.95$ & $0.944(1.067)$ & $0.383(0.958)$ &  & $%
0.946(1.047)$ & $0.422(0.948)$ &  & \\
&  & $0.99$ & $0.994(1.373)$ & $0.821(1.232)$ &  & $0.996(1.347)$ & $%
0.845(1.219)$ &  &\\ \hline
& \multirow{2}{*}{800} & $0.95$ & $0.940(0.941)$ & $0.296(0.845)$ &  & $%
0.944(0.931)$ & $0.296(0.843)$ &  &  \\
&  & $0.99$ & $0.997(1.207)$ & $0.752(1.084)$ &  & $0.999(1.193)$ & $%
0.776(1.080)$ &  & \\ \hline
& \multirow{2}{*}{$n$} & \multirow{2}{*}{$1-\alpha$} & \multicolumn{2}{c}{$%
\mbox{Case 3}$} &  & \multicolumn{2}{c}{$\mbox{Case 4}$} &  & \\
\cline{4-5}\cline{7-8}
&  & \multirow{2}{*} & $\mbox{SCB}$ & $\mbox{SCB-CC}$ &  & $\mbox{SCB}$ & $%
\mbox{SCB-CC}$ &  & \\ \hline
& \multirow{2}{*}{400} & $0.95$ & $0.922(1.103)$ & $0.582(1.019)$ &  & $%
0.933(1.150)$ & $0.595(1.045)$ &  &  \\
&  & $0.99$ & $0.993(1.425)$ & $0.899(1.316)$ &  & $0.995(1.481)$ & $%
0.905(1.346)$ &  &  \\ \hline
& \multirow{2}{*}{600} & $0.95$ & $0.944(0.947) $ & $0.511(0.873)$ &  & $%
0.946(0.987)$ & 0.473(0.894) &  &  \\
&  & $0.99$ & $0.997(1.214) $ & $0.881(1.119)$ &  & $0.997(1.262)$ &
0.881(1.143) &  & \\ \hline
& \multirow{2}{*}{800} & $0.95$ & $0.944(0.838) $ & $0.405(0.776)$ &  & $%
0.943(0.876)$ & $0.357(0.795)$ &  & \\
&  & $0.99$ & $0.993(1.071) $ & $0.846(0.991)$ &  & $0.999(1.116)$ & $%
0.805(1.013)$ &  &  \\ \hline\hline
\end{tabular}%
\end{center}
\end{table}

\begin{table}[tbp]
\caption{ Empirical coverage frequencies of the proposed SCB in (\protect\ref%
{EQ:feasible_SCB}) and the SCB in the complete case (SCB-CC) from 1000
replications and their corresponding average lengths (inside parentheses)
under using logistic regression to fit the underlying truncated logistic
selection probability with parameters $(\protect\alpha_0,\protect\alpha%
_1)=(0.2,0.6)$. }
\label{TAB:coverage_misprobit_05_02}
\begin{center}
\begin{tabular}{ccccccccccccc}
\hline\hline
& \multirow{2}{*}{$n$} & \multirow{2}{*}{$1-\alpha$} & \multicolumn{2}{c}{$%
\mbox{Case 1}$} &  & \multicolumn{2}{c}{$\mbox{Case 2}$} &  &   \\
\cline{4-5}\cline{7-8}
& \multirow{2}{*} &  & $\mbox{SCB}$ & $\mbox{SCB-CC}$ &  & $\mbox{SCB}$ & $%
\mbox{SCB-CC}$ &  &   \\ \hline
& \multirow{2}{*}{400} & $0.95$ & $0.901(1.262)$ & $0.395(1.107)$ &  & $%
0.909(1.222)$ & $0.473(1.102)$ &  &   \\
&  & $0.99$ & $0.991(1.634)$ & $0.782(1.434)$ &  & $0.992(1.582)$ & $%
0.859(1.426)$ &  &   \\ \hline
& \multirow{2}{*}{600} & $0.95$ & $0.924(1.073)$ & $0.237(0.939)$ &  & $%
0.924(1.030)$ & $0.329(0.925)$ &  &   \\
&  & $0.99$ & $0.993(1.381)$ & $0.713(1.208)$ &  & $0.994(1.327)$ & $%
0.803(1.191)$ &  &   \\ \hline
& \multirow{2}{*}{800} & $0.95$ & $0.923(0.948)$ & $0.160(0.831)$ &  & $%
0.933(0.917)$ & $0.220(0.825)$ &  &  \\
&  & $0.99$ & $0.995(1.216)$ & $0.599(1.066)$ &  & $0.996(1.175)$ & $%
0.707(1.057)$ &  &   \\ \hline
& \multirow{2}{*}{$n$} & \multirow{2}{*}{$1-\alpha$} & \multicolumn{2}{c}{$%
\mbox{Case 3}$} &  & \multicolumn{2}{c}{$\mbox{Case 4}$} &  &  \\
\cline{4-5}\cline{7-8}
&  & \multirow{2}{*} & $\mbox{SCB}$ & $\mbox{SCB-CC}$ &  & $\mbox{SCB}$ & $%
\mbox{SCB-CC}$ &  &    \\ \hline
& \multirow{2}{*}{400} & $0.95$ & $0.921(1.037)$ & $0.654(0.986)$ &  & $%
0.927(1.080)$ & $0.644(1.010)$ &  &  \\
&  & $0.99$ & $0.994(1.340)$ & $0.935(1.274)$ &  & $0.993(1.391)$ & $%
0.920(1.300)$ &  & \\ \hline
& \multirow{2}{*}{600} & $0.95$ & $0.937(0.885) $ & $0.591(0.841)$ &  & $%
0.944(0.926)$ & 0.556(0.864) &  &  \\
&  & $0.99$ & $0.995(1.136) $ & $0.930(1.079)$ &  & $0.998(1.184)$ &
0.904(1.105) &  &  \\ \hline
& \multirow{2}{*}{800} & $0.95$ & $0.932(0.786) $ & $0.512(0.748)$ &  & $%
0.942(0.823)$ & $0.435(0.770)$ & & \\
&  & $0.99$ & $0.994(1.004) $ & $0.884(0.956)$ &  & $0.997(1.048)$ & $%
0.858(0.980)$ &  &   \\ \hline\hline
\end{tabular}%
\end{center}
\end{table}

\newpage

\begin{figure}[tbp]
\center
\includegraphics[width=2.4in,height=2.2in]{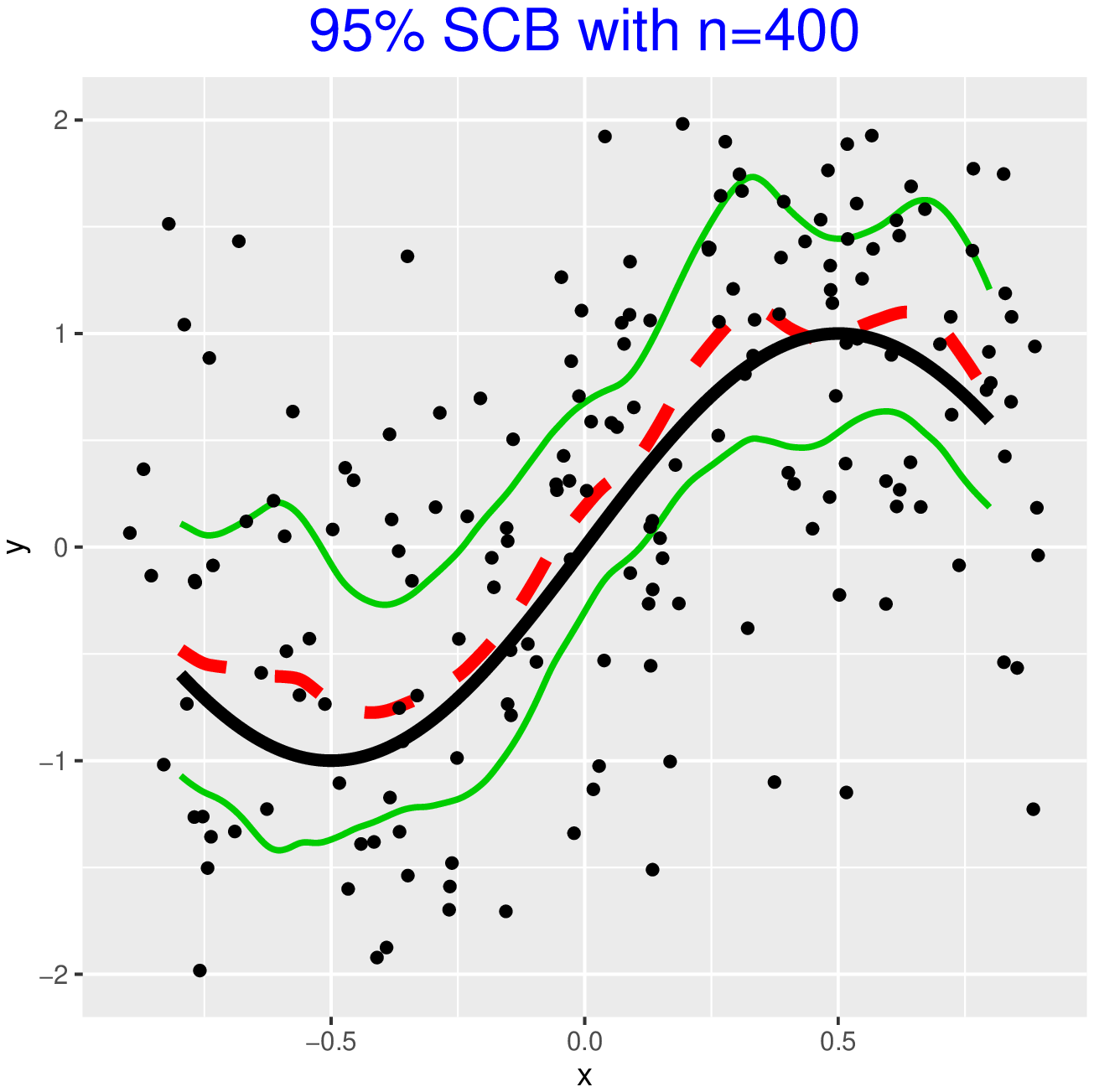} %
\includegraphics[width=2.4in,height=2.2in]{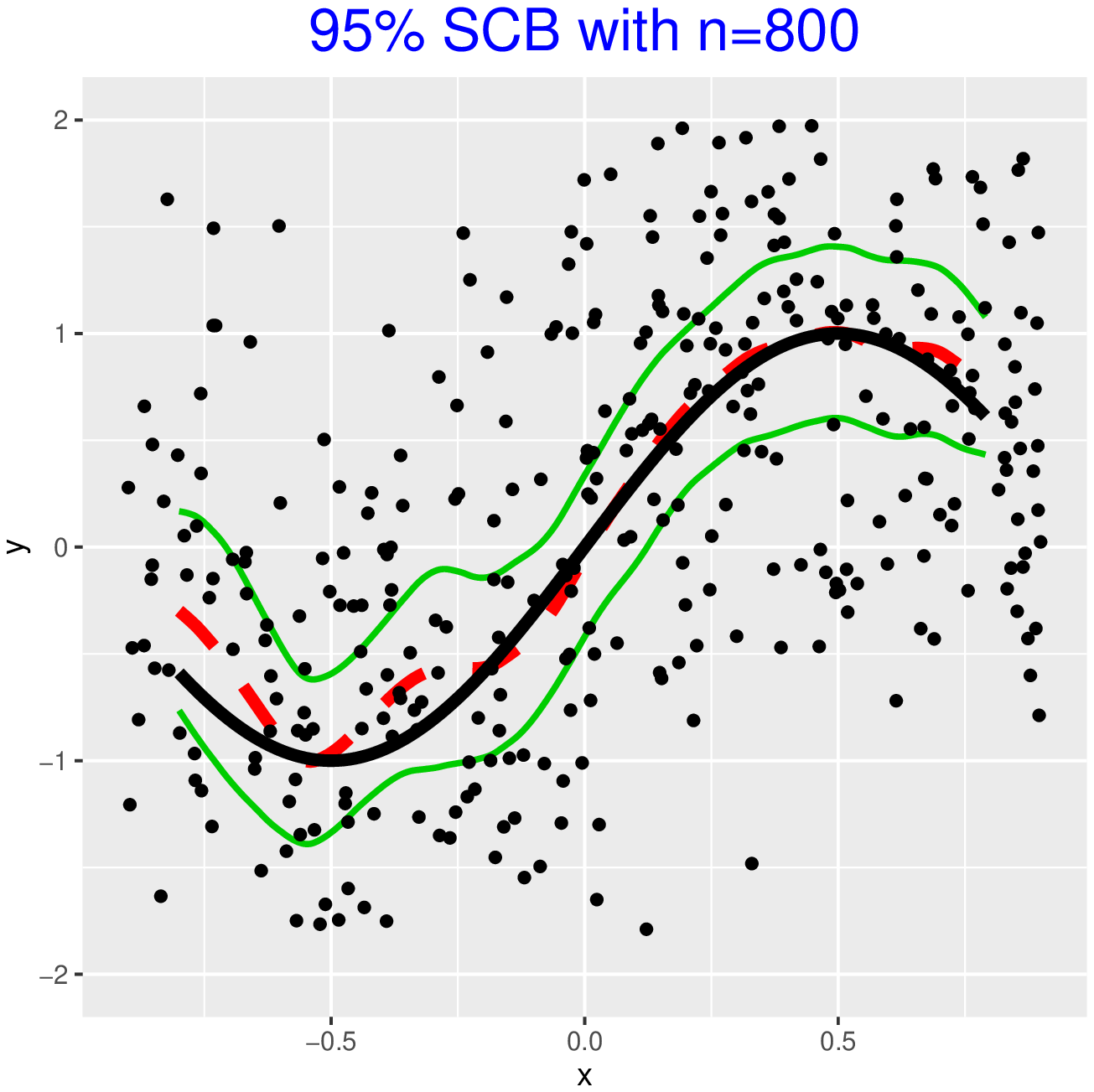}
\caption{Plots of the true mean function $m(x)$(thick solid line), the
weighted local linear estimate $\hat{m}(x,\hat{\protect\pi})$ (dashed line)
and the $95\%$ SCB (solid line) for Case 1 under the selection probability
model (i) with parameters $(\protect\alpha_0,\protect\alpha_1)=(0.2,0.6)$
(about $45\%$ missing). }
\label{Fig:plot_SCB_logit0206_case2}
\end{figure}

\begin{figure}[tbp]
\center
\includegraphics[width=2.4in,height=2.2in]{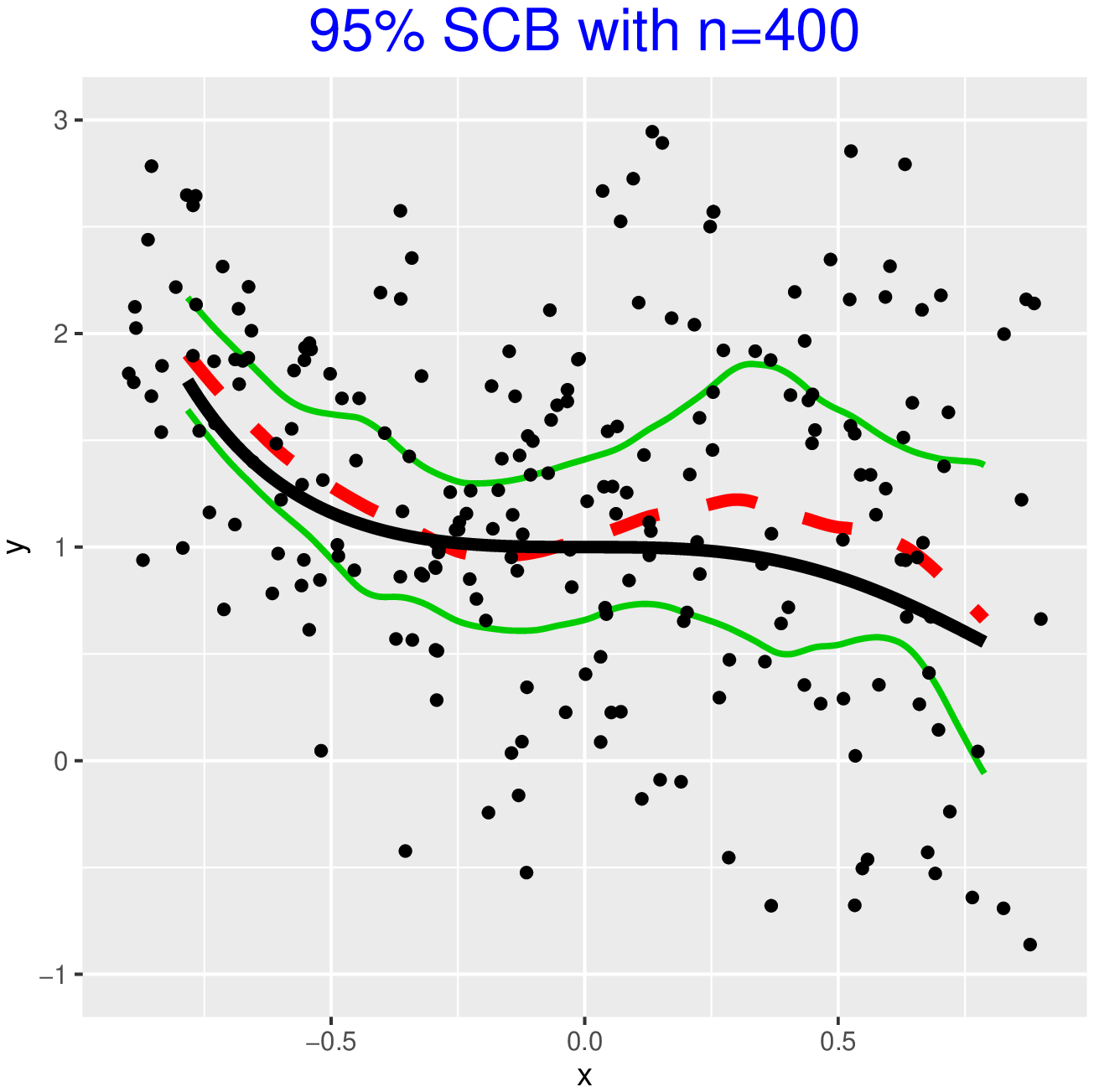} %
\includegraphics[width=2.4in,height=2.2in]{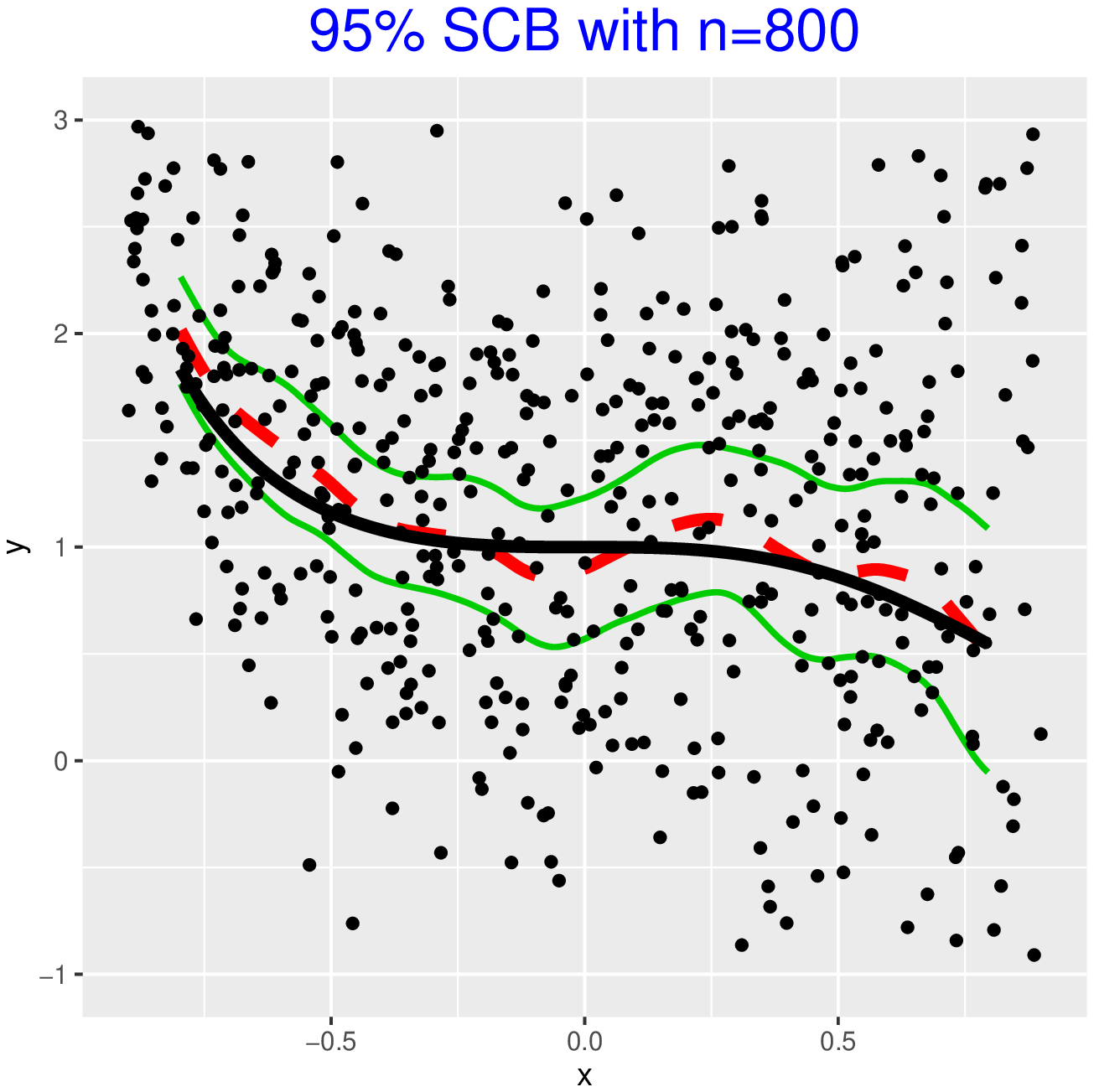}
\caption{Plots of the true mean function $m(x)$ (thick solid line), the
weighted local linear estimate $\hat{m}(x,\hat{\protect\pi})$ (dashed line)
and the $95\%$ SCB (solid line) for Case 4 under the selection probability
model (i) with parameters $(\protect\alpha_0,\protect\alpha_1)=(0.2,0.6)$
(about $31\%$ missing). }
\label{Fig:plot_SCB_logit0206_case4}
\end{figure}

\begin{figure}[tbp]
\center
\includegraphics[width=2.4in,height=2.2in]{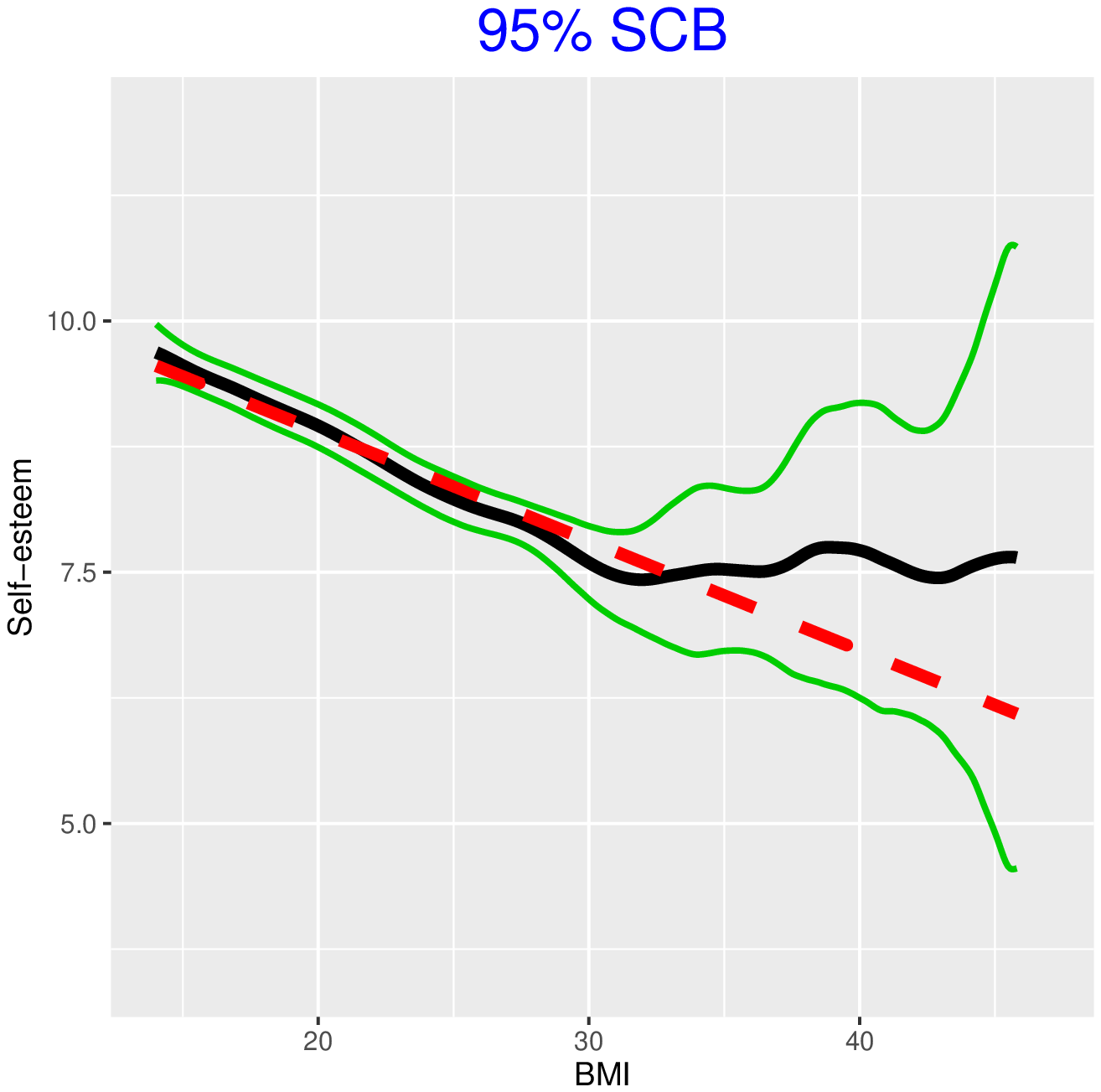} %
\includegraphics[width=2.4in,height=2.2in]{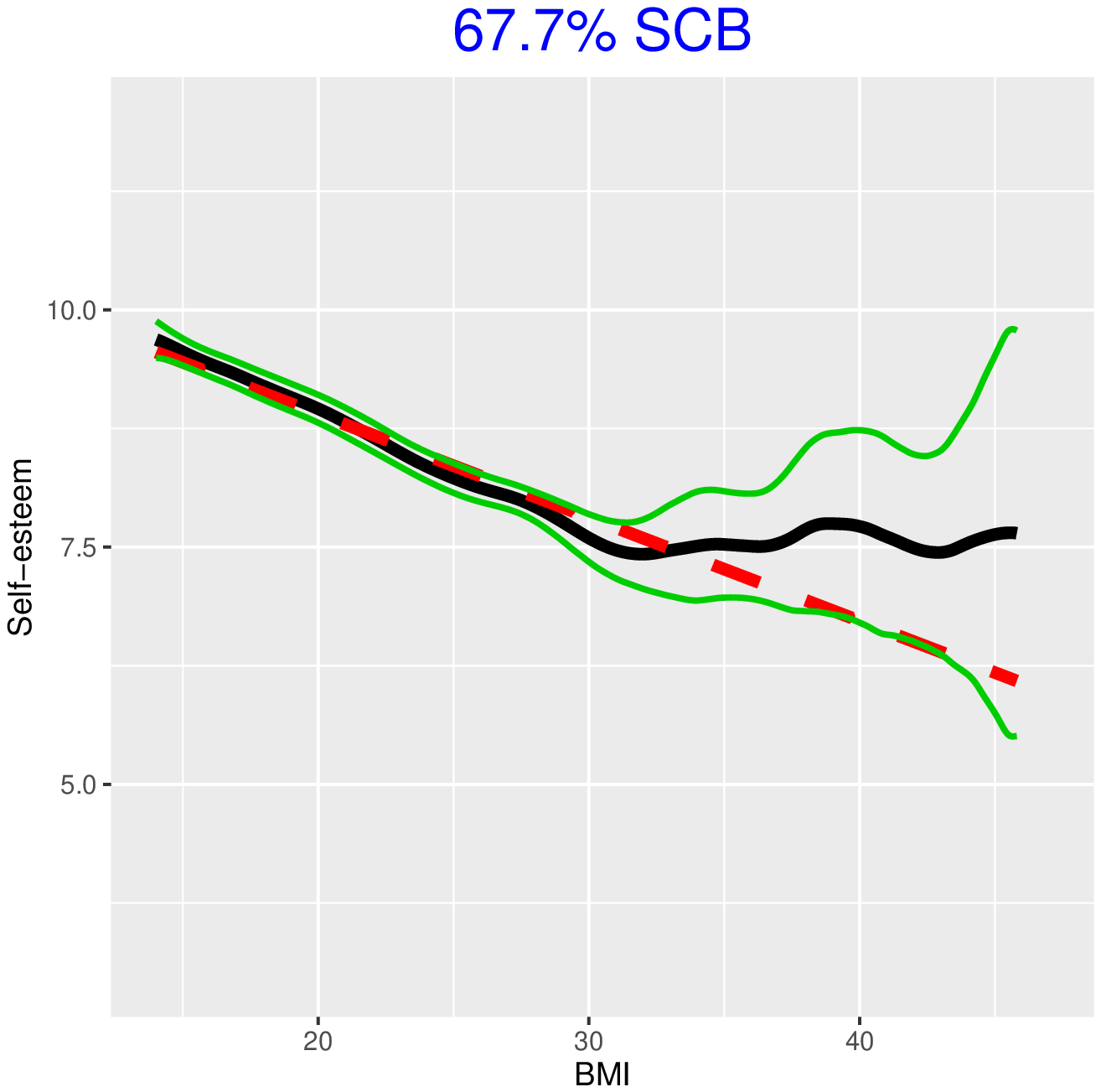}
\caption{Plots of the weighted local linear estimate $\hat{m}(x,\hat{\protect%
\pi})$ (thick solid line), the $95\%$ and $67.7\%$ SCBs (solid lines), and
the null hypothesis weighted linear regression curve (dashed line) for the
youth student survey data collected from white female students. }
\label{Fig:plot_SCB_real_data}
\end{figure}

\end{document}